\documentclass[journal,draftcls,onecolumn,12pt,twoside]{IEEEtranTCOM}
\normalsize

\usepackage{balance}
\usepackage[ruled]{algorithm2e}
\usepackage{amssymb}
\usepackage{mathbbol}
\usepackage{stmaryrd}
\usepackage{cite}
\usepackage{color}		
\usepackage{multirow}
\usepackage{subfigure}
\usepackage{graphicx,times,amsmath}
\usepackage{epstopdf}	
\usepackage{pifont}
\usepackage{indentfirst}
\usepackage{CJK}
\usepackage{amsfonts}
\usepackage{txfonts}

\usepackage{theorem}
\usepackage{float}
\usepackage{microtype}
\newtheorem{lemma}{Lemma}
\newtheorem{proposition}{Proposition}

\ifCLASSINFOpdf
\else
\fi

\begin{document}
\title{Joint Energy Harvest and Information Transfer for Energy Beamforming in Backscatter Multiuser Networks}
\author{Wenyuan~Ma,
	Wei~Wang,~\IEEEmembership{Senior Member,~IEEE},
	and~Tao~Jiang,~\IEEEmembership{Fellow,~IEEE}%
\thanks{Part of this work has been presented at IEEE GLOBECOM~\cite{wen2019}.
	
	W. Ma, W. Wang and T. Jiang are with the School of Electronic Information and Communications, Huazhong University of Science and Technology. E-mail: \{wenyuan\_ma,weiwangw,taojiang\}@hust.edu.cn.}}

%

\maketitle

\begin{abstract}
Wirelessly powered backscatter communication (WPBC) has been identified as a promising technology for low-power communication systems, which can reap the benefits of energy beamforming to improve energy transfer efficiency. However, existing studies on energy beamforming fail to simultaneously take energy supply and information transfer in WPBC into account. This paper takes the first step to fill this gap, by considering the restrictive relationship between the energy harvesting rate and achievable rate with estimated backscatter channel state information (BS-CSI). To ensure reliable communication and user fairness, we formulate the energy beamforming design as a max-min optimization problem by maximizing the minimum achievable rate for all backscatter tags subject to the energy constraint. We derive the closed-form expression of the energy harvesting rate, as well as the lower bound of the achievable rate for maximum-ratio combining (MRC) and zero-forcing (ZF) receivers. Our numerical results indicate that our scheme significantly outperforms state-of-the-art energy beamforming schemes. Additionally, the achievable rate of our scheme approaches more than $90\%$ of the rate limit achieved via beamforming with perfect CSI for both receivers.
\end{abstract}


%
\IEEEpeerreviewmaketitle

\section{Introduction}
\IEEEPARstart{W}{irelessly} powered backscatter communication (WPBC) has recently emerged as a promising technology for pervasive sustainable Internet-of-Things (IoT) applications, such as smart city, connected health, and wildlife tracking~\cite{8253544,7876867,8527670,8093703}. Unlike conventional wireless communication that relies on power-hungry and high-cost radio frequency (RF) components for data transmission, WPBC technology exploits lightweight and batteryless backscatter tags to transfer information by intermittently harvesting and reflecting ambient wireless signals from TV, cellular, or WiFi transmissions~\cite{2013ambient,2016Passivewifi,Authenticating,2018shieldscatter}. However, due to the small amount of energy harvested by a backscatter tag and double power attenuation at the receiver, WPBC is limited to short communication ranges and low bit rates, impeding its widespread applications~\cite{6942226,7937935,8368232}. Therefore, to promote the application of WPBC systems, it is crucial to improve the energy and information transfer efficiency by developing novel designs on the active transceiver, leaving backscatter tags as simple as possible.
	
With the potential of increasing both wireless power and energy efficiency, energy beamforming that forms sharp energy beams towards target users according to instantaneous channel state information (CSI), has attracted much attention in both academia and industry~\cite{zhang2013mimo,yang2015throughput,6954434,2017powerbased,2018rssi,psomas2019energy,chen2018beamforming}. Growing attempts have been devoted to exploring the merits of energy beamforming in WPBC~\cite{long2017transmit,gong2018backscatter,2019ReciprocityBased,2019sumthroughput,2019Multi-Tag,yang2015multi} under different CSI conditions. From an information transfer perspective, Long \textit{et al.}~\cite{long2017transmit} propose a transmit beamforming scheme to maximize the sum rate under the assumption of perfect CSI availability in a WPBC system, where a cooperative receiver decodes information from both a primary transmitter and a secondary backscatter tag. Gong \textit{et al.}~\cite{gong2018backscatter} investigate a throughput maximization problem by using energy beamforming and the relay strategy in a relay WPBC system under perfect channel information and uncertain channel information. Recently, for a point-to-point monostatic WPBC system consisting of a multiple-input-multiple-output (MIMO) transmitter and a backscatter tag, Mishra \textit{et al.}~\cite{2019ReciprocityBased} present an energy beamforming scheme that obtains the CSI using a least-square (LS) estimator under the channel reciprocity to maximize the average backscattered signal-to-noise ratio (SNR) for efficient information transfer. Furthermore, Mishra \textit{et al.} extend it to WPBC systems with multiple backscatter tags for maximizing the achievable sum-backscattered throughput~\cite{2019sumthroughput} and realizing throughput fairness~\cite{2019Multi-Tag}, respectively. From an energy transfer perspective, Yang \textit{et al.}~\cite{yang2015multi} investigate energy beamforming designs to improve wireless energy transfer (WET) efficiency from the reader to multiple backscatter tags via limited channel estimations with a single receive antenna.

Despite the fact that much has been understood through these studies, their beamforming designs fail to take the fundamental restrictive relationship between energy supply and information transfer into account~\cite{8476597}. These studies either only focus on optimizing the transmit beamforming to maximize the achievable rate of WPBC systems~\cite{long2017transmit,gong2018backscatter,2019ReciprocityBased,2019sumthroughput,2019Multi-Tag}, or design energy beamforming just to maximize a total utility of the harvested energy~\cite{yang2015multi}. However, there are different optimal beams for the achievable rate and energy harvesting rate. The reason lies in the fact that the energy harvesting rate at the backscatter tag only depends on the forward channel (i.e., transmitter-to-tag), while the achievable rate at the reader depends on the backscatter channel (i.e., transmitter-to-tag-to-receiver). As a consequence, energy beamforming that only maximizes harvested energy cannot guarantee the achievable rate to retain reliable communications. Likewise, energy beamforming that only maximizes the achievable rate cannot ensure the energy supply.
\begin{figure}[!t]
	\centering
	\includegraphics[width=0.73\linewidth]{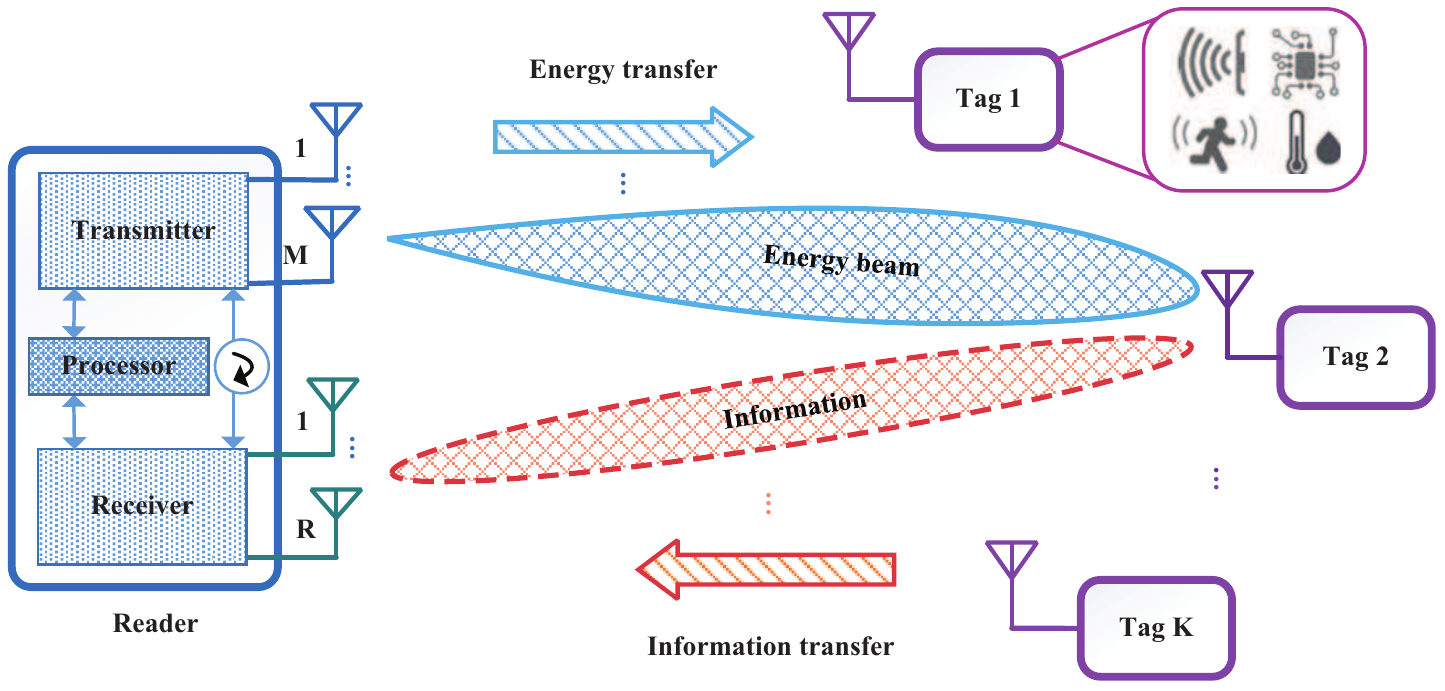}
	\caption{A WPBC network with energy beamforming.}
	\label{fig:figure1}
\end{figure}

To overcome the above limitation, this paper takes the first step to investigate the energy beamforming design that involves the restrictive relationship between energy supply and information transfer. As shown in Fig.~\ref{fig:figure1}, this paper considers a general monostatic WPBC network where a reader with multiple antennas transmits energy to multiple single-antenna backscatter tags, and tags transmit information to the reader by reflecting incident carrier signals. One critical issue in implementing the energy beamforming in WPBC systems is to obtain the instantaneous forward channel state information (F-CSI) for the bistatic antenna reader, which, however, is rather challenging due to the closed-loop propagation and power-limited tags. We tackle this challenge by estimating the backscatter channel state information (BS-CSI) using both LS and minimum mean squared error (MMSE) approaches at the reader to perform energy beamforming instead of F-CSI. In particular, this pushes the computational complexity and energy consumption from backscatter tags to the reader, leaving tags as simple as possible.

Besides, to ensure reliable communication and user fairness, we formulate the beamforming design as a max-min optimization problem by maximizing the minimum achievable rate for all backscatter tags subject to tags' power consumption constraints. It is particularly hard to quantify the essential metrics (e.g., the energy harvesting rate and achievable rate), which require the evaluation of expectation with respect to the distribution of F-CSI. However, the distribution of the actual forward channel is unknown. To address this predicament, we leverage Bayesian philosophy to analytically derive the distribution of the actual forward channel conditioned on the estimated BS-CSI. Furthermore, we obtain closed-form expressions for the energy harvesting rate, as well as the lower bound of the ergodic achievable rate for maximum-ratio combining (MRC) and zero-forcing (ZF) receivers.  

The major contributions of this paper are summarized as follows.
\begin{itemize}
	\item 	We propose a beamforming scheme for energy supply and information transfer in WPBC using the estimated BS-CSI instead of commonly used F-CSI. To the best of our knowledge, this is the first work that considers the restrictive relationship between the energy harvesting rate and the achievable data rate in WPBC.
\end{itemize}
\begin{itemize}
    \item We perform channel estimation based on LS and MMSE criteria for a bistatic antenna reader to estimate the backscatter channel matrix, and further analyze the conditional distributions of forward and backward channels (i.e., tag-to-receiver) under each other's influence.
\end{itemize}
\begin{itemize}
	\item 	We obtain the analytical expressions of the energy harvesting rate and lower bound of the ergodic achievable rate using low-complexity and asymptotically-optimal linear MRC and ZF receivers with the estimated BS-CSI, considering the effect of unknown forward and backward channels.
\end{itemize}
\begin{itemize}
	\item 	Our numerical results indicate that the minimum achievable rate of the proposed scheme is significantly higher than that of previous energy beamforming schemes, while slightly lower than that obtained by energy beamforming with perfect CSI. 
\end{itemize}

The rest of this paper is organized as follows. Section~\ref{sec:model} establishes the system model. Section~\ref{sec:estimator} describes the backscatter channel estimation. Section~\ref{sec:design} elaborates on the energy beamforming design for energy supply and information transfer.  Next, Section~\ref{sec:analysis} analyzes the harvested energy and achievable rate of the proposed scheme. Section~\ref{sec:evaluation} shows the numerical results. Finally, Section~\ref{sec:Extensions} discusses some future work and Section~\ref{sec:conclusion} concludes our paper, respectively.

\section{System Model}\label{sec:model}
In this section, we first present the system architecture, slot structure, and backscatter channel model. Then, we describe the energy transfer phase and backscatter phase in the proposed energy beamforming scheme for backscatter communications.
 
\begin{table}[h]
	\renewcommand\arraystretch{0.62}
	\centering
	\caption{Symbols and notations}\label{tab:tab1}
	\begin{tabular}{|c|l|c|l|}
		\hline
		\textbf{Symbol} & \textbf{Definition}  & \textbf{Symbol} & \textbf{Definition} \\\hline
		$T$   & Duration of each time block & $\alpha$  & Duration of channel estimation \\\hline
		\multirow{2}{*}{$h_{mk}^{f}$}  & Forward channel coefficient between the $m$-th   & \multirow{2}{*}{$h_{kr}^{b}$}  & Backward channel coefficient between the $k$-th \\ 
		&  smit  antenna of the reader and tag $k$ & & tag and the $r$-th receive antenna of the reader \\\hline
		\multirow{4}{*}{$h_{mkr}$} & Backscatter channel coefficient between the $m$-th& \multirow{4}{*}{$\beta_{k}$} & Path loss of the channel between the \\
		&  transmit antenna and the $r$-th receive antenna& & reader and tag $k$\\
		&   of the reader while passing through tag $k$ & & \\\hline
		$\textbf{H}_{k}$ & Backscatter channel matrix of tag $k$ & $\boldsymbol{\Phi}$ & Beamformer vector \\\hline
		$u$ & Transmitted carrier signal from reader & $p$ & Transmit power of the carrier signal \\\hline
		\multirow{2}{*}{$\delta_{k}$} & Power reflection coefficient of tag $k$, & \multirow{2}{*}{$p_{ce}$} & Power of each antenna transmitting a pilot seq- \\
		& $\delta_{k}\in{\mathbb{C}} \left( 0\leqslant\delta_{k}\leqslant1 \right)$ & & uence in CE slot\\\hline
		$P_{Ik}$ & Incident signal power of tag $k$ & $ P_{Ek}$ &  Instantaneous energy harvesting rate \\\hline
		$w$ & Average transmit power for each time block  & $\eta$ & Rectifier efficiency, $\eta\in\left(0,1\right]$ \\\hline
		$s_{k}$ & Backscatter signal of tag $k$ &  $\textbf{Q}$ & A linear detector \\\hline
		\multirow{2}{*}{$R_{k}$} & Achievable rate of the information transmission  & \multirow{2}{*}{$D$} & The length of pilot sequence \\
		& from tag $k$ & &\\\hline
		\multirow{2}{*}{$\sigma_{\tilde{e},k}^2$} & Backscatter channel estimation error variance   &  \multirow{2}{*}{$\sigma_{e,kr}^2(h_{kr}^{b})$} & Estimation error variance of forward channel  \\
		& of tag $k$ & & conditioned on backward channel parameter $h_{kr}^{b}$\\\hline
		\multirow{2}{*}{$\sigma_{\epsilon,mi}^2(h_{mi}^{f})$} & Estimation error variance of backward channel  & \multirow{2}{*}{$\zeta_{k}$} & Energy allocation weights of the beamformer,\\
		& conditioned on forward channel parameter $h_{mi}^{f}$ & &  $\zeta_{k}\in{[0,1]}$ \\\hline
		$\rho$ & Circuit power consumption rate in backscattering & $\tau$ &  Minimum of  $|h_{mi}^f|^2$,  $\tau>0$\\
		\hline
	\end{tabular}
\end{table}

\subsection{System Architecture and Slot Structure}
As shown in Fig.~\ref{fig:figure1}, we consider a general WPBC system comprising of $K$ single-antenna backscatter tags and one bistatic antenna reader equipped with $M$ transmit antennas and $R$ receive antennas. First, this reader transmits carrier signals to the backscatter tags. Then, each tag harvests energy to power up its circuit and conveys information to the reader by reflecting incident carrier signals. Finally, the reader receives the backscattered signals and decodes messages from them. 

We assume transmission over time blocks. The important symbols used in this paper are given in TABLE I. The duration of each time block $T$ is shorter than the coherent interval, each of which consists of a channel estimation (CE) slot and a simultaneous wireless energy and information transfer slot. Specifically, the reader first sends pilots and leverages backscattered signals to estimate the BS-CSI associated with each backscatter tag. Subsequently, the reader performs energy beamforming to transmit wireless energy and provide carrier signals for all tags. Simultaneously, tags transfer information to the reader. 

\subsection{Channel Model}

Let $\textbf{h}_{k}^{f}=\left[h_{1k}^{f}, h_{2k}^{f}, ... ,h_{Mk}^{f} \right]^T\in\mathbb{C}^{M\times1}$ and $\textbf{h}_{k}^{b}=\left[h_{k1}^{b}, h_{k2}^{b}, ... , h_{kR}^{b} \right]^T\in\mathbb{C}^{R\times1}$ denote the forward and backward channel vector, respectively. With longer coverage range of WPBC networks and isotropic wave propagation, channels are characterized by rich scattering in a complex environment and hence considered no dominant line of sight (LoS)  path between the reader and any tags. To formulate the links that no dominant propagation along a line of sight, the Rayleigh fading model is a statistical model constructed based on empirical studies and is well-accepted in the research community. Thus, we consider a Rayleigh-fading channel just like many existing studies about backscatter communication, e.g.,~\cite{2019Multi-Tag,18Inference,19Constellation}. Thus, we assume that $h_{mk}^{f}$ and $h_{kr}^{b}$ are independent and identically distributed (i.i.d.) as $\mathcal{CN}\left(0, \beta_{k }\right)$, where $\beta_{k}$ is constant over the coherent interval and known at the reader. Let $\textbf{H}_{k}=\textbf{h}_{k}^{b}\left(\textbf{h}_{k}^{f}\right)^{T}$ denote the $R\times M$ backscatter channel matrix of tag $k$. Its element $[\textbf{H}_{k}]_{mr}=h_{mkr}=h_{mk}^{f}h_{kr}^{b}$ is the backscatter channel coefficient, which is a combination of the channel from the $m$-th transmit antenna to tag $k$ and the channel from tag $k$ to the $r$-th receive antenna of the reader.

\subsection{Energy Transfer}
In the WPBC system, the reader transmits a carrier signal to $K$ tags concurrently. To improve the energy transmission efficiency, we use energy beamforming and assign a beamformer vector $\boldsymbol{\Phi}$ to the carrier signal $u\in\mathbb{C}$ with $\mathbb{E} \left\lbrace \left| u \right|^2  \right\rbrace=p$ with an average transmit power of $w$. Thus, we can obtain $p=\frac{wT -\alpha p_{ce}}{T-\alpha} $,  where $p_{ce}$ represents the power of each antenna transmitting a pilot sequence in CE slot. Thereby, the signal emitted by the reader is given as $u\boldsymbol{\Phi}$, and the received signal at tag $k$ is given by
    \begin{equation}
    b_{k} =u\boldsymbol{\Phi}^T\textbf{h}_{k}^{f} + n_{k},
    \end{equation}
where $n_{k}\sim \mathcal{CN}\left( 0, \sigma_{k}^2\right)$ denotes the noise.

Note that the noise energy cannot be harvested~\cite{yang2015multi}. Then, the incident signal power of tag $k$ can be represented as 
    \begin{equation}\label{sec2:incident signal power}
    P_{Ik}=\mathbb{E} \left\lbrace \left| u \right|^2 |\boldsymbol{\Phi}^T\textbf{h}_{k}^{f}|^2  \right\rbrace
    =p\mathbb E\left\lbrace|\boldsymbol{\Phi}^T\textbf{h}_{k}^{f}|^2 \right\rbrace.
    \end{equation}
%

When the tag is activated to reflect the incoming signal, a fraction of the incident signal power, denoted as $(1-\delta_{k})P_{Ik}$, is rectified for conversion from RF signal to direct current (DC)~\cite{lu2018wireless}. Otherwise, the whole incident signal power is rectified to DC, i.e., $\delta_{k}=0$. However, due to hardware limitations, RF signals cannot be fully converted to DC. There are various models to describe the relationship between the harvested energy and input power. To simplify the analysis, we assume that the harvested energy scales linearly with the input power~\cite{khan2018optimization} with the rectifier efficiency $\eta$. Thus, the instantaneous energy harvesting rate for an active tag, denote by $P_{Ek}$, is the proportional of the incident signal power, i.e.,
	\begin{equation}\label{energy harvesting rate}
	P_{Ek}=\eta(1-\delta_{k})P_{Ik}.
	\end{equation} 

\vspace{-1cm}
\subsection{Information Transfer}
When the instantaneous energy harvesting rate exceeds the circuit power consumption rate, tags reflect and modulate the incident carrier signal. Then, the received signal vector from tag $k$ can be written as
    \begin{equation}
    \begin{aligned}
    \textbf{y}_{k}& =\sqrt{\delta_{k}}\textbf{h}_{k}^{b}b_{k}s_{k}+\textbf{z}_{k}\\   &=\sqrt{\delta_{k}}us_{k}\textbf{h}_{k}^{b}\left(\textbf{h}_{k}^{f}\right)^{T}\boldsymbol{\Phi} + \textbf{v}_{k}.
    \end{aligned} 
    \end{equation}
We assume $s_{k}\sim\mathcal{CN}\left( 0, 1 \right) $, and the noise $\textbf{v}_{k}\sim\mathcal{CN}\left( \textbf{0}, \boldsymbol{\Sigma} \right) $~\cite{long2017transmit}.

Hence, the received signal vector at the reader can be represented by
	\begin{equation}
	\begin{aligned}
	\textbf{y} = \textbf{H}_{direct} u\boldsymbol{\Phi}+\sum_{k=1}^{K} \textbf{y}_{k},
	\end{aligned} 
	\end{equation}
	where $\textbf{H}_{direct}$ is the direct channel from the reader's transmit antenna array to its receive antenna array. The first term in the above equation is self-interference stemmed from the direct path between the reader's transmit antennas and the receive antennas. Furthermore, since tags' signal transmission depends on the carrier signal sent by the reader, self-interference is inevitable in backscatter communication. Thus, to decode the tag's backscatter signal, we first eliminate the self-interference by employing self-interference cancellation techniques~\cite{Backfi,15SecuringRFIDs}. We briefly introduce the general procedure of self-interference cancellation using components such as cancellation filters, couplers, etc., but refer~\cite{Backfi} for a complete description. First, the reader estimates the channel $\textbf{H}_{direct}$. Then, the estimated channel distortion is applied to recreate the self-interference accurately with the known transmit signal from the reader's transmit antenna. Finally, the reader can subtract the self-interference from the received signal, and then the received signal vector at the reader from $K$ tags can be represented by
    \begin{equation}
    \begin{aligned}\label{RSV_IT}
    \textbf{y} = \sum_{k=1}^{K} \textbf{y}_{k} &=\textbf{H}^{b}\left[\left(\left(\textbf{H}^{f}\right)^{T}u\boldsymbol{\Phi}\right)\circ\left(\boldsymbol{\Lambda}\textbf{s}\right)\right]+\boldsymbol{\mu} =\textbf{H}^{b}\textbf{x}+\boldsymbol{\mu},
    \end{aligned} 
    \end{equation}
where $\textbf{H}^{b}=[\textbf{h}_{1}^{b}, \textbf{h}_{2}^{b}, ... , \textbf{h}_{K}^{b}]$, $\textbf{H}^{f}=[\textbf{h}_{1}^{f}, \textbf{h}_{2}^{f}, ... , \textbf{h}_{K}^{f}]$, $\boldsymbol{\Lambda}=diag\{\sqrt{\delta_{1}}, \sqrt{\delta_{2}}, ... , \sqrt{\delta_{K}}\}$, $\textbf{s}=[s_{1}, s_{2}, ... , s_{K}]^T$, $\textbf{x}=[\sqrt{p_{1}}s_{1}, \sqrt{p_{2}}s_{2}, ... , \sqrt{p_{K}}s_{K}]^T$. Besides, $p_{k} = \delta_{k} P_{Ik}$ denotes the reflect power, and $\boldsymbol{\mu}$ is a vector of additive white Gaussian noise. The element of noise vector $u_{r}\sim\mathcal{CN}\left( 0, \sigma^2 \right)$.

Then, the received signal vector can be spatially separated by using a linear detector $\textbf{Q}$ to detect the information from all tags, which is given by 
    \begin{equation}
    \begin{aligned}
    \textbf{r} &= \textbf{Q}^{H} \textbf{H}^{b} \textbf{x} + \textbf{Q}^{H} \boldsymbol{\mu}. 
    \end{aligned} 
    \end{equation}
Let $r_{k}$ and $x_{k}$ be the $k$-th element of the $K \times 1$ vectors $\textbf{r}$ and $\textbf{x}$, $\textbf{q}_{k}$ and $\textbf{h}_{k}^{b}$ the $k$-th column of the detector $\textbf{Q}$ and channel $\textbf{H}^{b}$. Therefore, after using the linear detector, the received signal associated with the $k$-th tag can be written as 
    \begin{equation}
    r_{k} = \textbf{q}_{k}^{H} \textbf{h}_{k}^{b} x_{k} + \sum_{i=1,i\neq k}^{K} \textbf{q}_{k}^{H} \textbf{h}_{i}^{b} x_{i} + \textbf{q}_{k}^{H} \boldsymbol{\mu}.
    \end{equation}
The ergodic achievable rate of the information transmission from tag $k$ is then given by
    \begin{equation}\label{rate}
    R_{k} \triangleq \mathbb{E} \left\lbrace \log_{2}\left( 1 + \frac{ p_{k} |\textbf{q}_{k}^{H} \textbf{h}_{k}^{b}|^2 }{\sum_{i=1,i\neq k}^{K}  p_{i} |\textbf{q}_{k}^{H} \textbf{h}_{i}^{b}|^2  + |\textbf{q}_{k}^{H} \textbf{q}_{k}|\sigma^2} \right) \right\rbrace.  
    \end{equation}

We can observe from the above derivations that the CSI is a crucial prerequisite for energy beamforming. Therefore, it is essential to get accurate CSI to obtain the energy harvesting rate and the achievable rate.

\section{Backscatter Channel Estimation}\label{sec:estimator}
Since the CSI affects the performance of energy beamforming and linear detection of the multi-antenna reader, CE quality is critical. However, due to the closed-loop propagation and hardware-limited tags, it is infeasible for backscatter devices to estimate F-SCI and then feedback it to the reader. Thus, in this section, we estimate the BS-CSI, which pushes the CE complexity and power consumption to the reader. 
Furthermore, to consider the impact of the tradeoff between complexity and estimation performance on energy beamforming design, we analyze the backscatter CE methods under both LS and MMSE criteria.

In the backscatter CE slot, the reader estimates the channel matrix of each tag by controlling tags to reflect the pilot signals in sequence~\cite{yang2015multi}. When estimating the backscatter channel of the $k$-th tag, the reader transmits orthogonal pilot sequences $\textbf{G} \textbf{B}^{1/2}\in{\mathbb{C}^{D\times M}}$, where $\textbf{G}\in{\mathbb{C}^{D \times M}}$ satisfies $\textbf{G}^{H}\textbf{G}=\textbf{I}_{M}$, and $\textbf{B}\in{\mathbb{C}^{M \times M}}$ is a diagonal matrix with $\left[\textbf{B}\right]_{ij}=Dp_{ce}$. 
Then, $K$ tags reflect the pilot signals with their power reflection coefficient $\delta_{k}$ in sequence. 

The reader can eliminate the direct-link influence as described in Section~\ref{sec:model}.D. Thus, the received signal at the reader is given by
    \begin{equation}
    \begin{aligned}
    \textbf{Y}_{k}^{CE} &=\sqrt{\delta_{k}}\textbf{H}_{k}(\textbf{G} \textbf{B}^{1/2})^{T}+\textbf{N}_{k},
    \end{aligned}
    \end{equation} 
where $\textbf{N}_{k}$ is an $R\times D$ matrix while elements are i.i.d. and $\left[\textbf{N}_{k}\right]_{mr} = n_{k,mr} \sim \mathcal{CN}\left( 0, \sigma^2\right)$.

\underline{LS Channel Estimation.}
First, we use the LS approach to estimate the backscatter channel.
Given $\textbf{Y}_{k}^{CE}$, the LS estimate of $\textbf{H}_{k}$ can be expressed as
    \begin{equation}
    \begin{aligned}
    \widehat{\textbf{H}}_{k}^{LS}&=\textbf{Y}_{k}^{CE} \delta_{k}^{-1/2}\textbf{G}^{*}\textbf{B}^{-1/2}=\textbf{H}_{k}+\textbf{N}_{k}\delta_{k}^{-1/2}\textbf{G}^{*}\textbf{B}^{-1/2}.
    \end{aligned}
    \end{equation}  
Denote the estimation error by $\widetilde{\textbf{E}}_{k}^{LS}\triangleq \widehat{\textbf{H}}_{k}^{LS}-\textbf{H}_{k}$, where $\tilde{e}_{k,mr}^{LS}$, $\hat{h}_{mkr}^{LS}$, and $h_{mkr}$ represent the $m$-th row, $r$-th column element of $\widetilde{\textbf{E}}_{k}^{LS}$, $\widehat{\textbf{H}}_{k}^{LS}$, and $\textbf{H}_{k}$, respectively. Moreover, $\widetilde{\textbf{E}}_{k}^{LS}$ is independent of $\widehat{\textbf{H}}_{k}^{LS}$. Thus, we can obtain $\tilde{e}_{k,mr}^{LS}\sim \mathcal{CN}\left( 0, (\sigma_{\tilde{e},k}^{LS})^2\right)$, where $(\sigma_{\tilde{e},k}^{LS})^2=\frac{\sigma^2}{Dp_{ce}\delta_{k}}$.

\underline{MMSE Channel Estimation.}
We now consider that the reader estimates the backscatter channel by using the MMSE estimator. The MMSE estimate result of $\textbf{H}_{k}$ can be expressed as
	\begin{equation}
	\begin{aligned}
	\widehat{\textbf{H}}_{k}^{MMSE}&=\textbf{Y}_{k}^{CE} \delta_{k}^{-1/2}\left(\textbf{G}^{*}\textbf{R}_{\textbf{H}_{k}}\textbf{B}\textbf{G}^{T}+\sigma^2R\textbf{I}\right)^{-1}\textbf{G}^{*}\textbf{B}^{1/2}\textbf{R}_{\textbf{H}_{k}}\\&\overset{(a)}{=}\textbf{Y}_{k}^{CE} \delta_{k}^{-1/2}\textbf{G}^{*}\left(\textbf{R}_{\textbf{H}_{k}}\textbf{B}+\sigma^2R\textbf{I}\right)^{-1}\textbf{B}^{1/2}\textbf{R}_{\textbf{H}_{k}}\\
	&=\textbf{Y}_{k}^{CE} \delta_{k}^{-1/2}\textbf{G}^{*}\left(\boldsymbol{\Psi}_{k}\textbf{B}+\sigma^2\textbf{I}\right)^{-1}\textbf{B}^{1/2}\boldsymbol{\Psi}_{k}, 
	\end{aligned} 
	\end{equation}
where $(a)$ is from the matrix inversion identity $(\textbf{I}+\textbf{A}\textbf{B})^{-1}\textbf{A}=\textbf{A}(\textbf{I}+\textbf{B}\textbf{A})^{-1}$, $\textbf{R}_{\textbf{H}_{k}}=\mathbb E \left[\textbf{H}_{k}^{H}\textbf{H}_{k}\right]=diag\{R\beta_{k}^2, R\beta_{k}^2, ..., R\beta_{k}^2\}=R\boldsymbol{\Psi}_{k}$. 
	
Denote the estimation error by $\widetilde{\textbf{E}}_{k}^{MMSE}\triangleq \widehat{\textbf{H}}_{k}^{MMSE}-\textbf{H}_{k}^{MMSE}$, where $ \tilde{e}_{k,mr}^{MMSE}$, $\hat{h}_{mkr}^{MMSE}$, and $h_{mkr}$ represent the $m$-th row , $r$-th column element of $\widetilde{\textbf{E}}_{k}^{MMSE}$, $\widehat{\textbf{H}}_{k}^{MMSE}$, and $\textbf{H}_{k} $, respectively. Besides, $\widetilde{\textbf{E}}_{k}^{MMSE}$ is independent of $\widehat{\textbf{H}}_{k}^{MMSE}$.
	
The performance of the MMSE estimator is characterized by the error matrix $\widetilde{\textbf{E}}_{k}^{MMSE}$ with zero mean and row covariance matrix
	\begin{equation}
    \begin{aligned}
    \textbf{R}_{\widetilde{\textbf{E}}_{k}^{MMSE}}&=\mathbb{E} \left[\widetilde{\textbf{E}}_{k}^{MMSE}(\widetilde{\textbf{E}}_{k}^{MMSE})^{H}\right] =\left(\textbf{R}_{\textbf{H}_{k}}^{-1}+\delta_{k}\textbf{B}\textbf{R}_{\textbf{N}_{k}}^{-1}\right)^{-1} =\frac{R\beta_{k}^2}{(1+\frac{\delta_{k}\beta_{k}^2Dp_{ce}}{\sigma^2})}\textbf{I}_{M}. 
    \end{aligned} 
    \end{equation}
Thus, we can obtain $\tilde{e}_{k,mr}^{MMSE}\sim \mathcal{CN}\left( 0, (\sigma_{\tilde{e},k}^{MMSE})^2\right)$, where $(\sigma_{\tilde{e},k}^{MMSE})^2=\frac{\beta_{k}^2}{(1+\frac{\delta_{k}\beta_{k}^2Dp_{ce}}{\sigma^2})}$.

We can estimate BS-CSI using LS and MMSE estimators. However, energy beamforming needs to obtain F-CSI. Therefore, we use the backward channel state information (B-CSI) as an unknown parameter to obtain the desired F-CSI through BS-CSI. Mathematically, since the elements of the backward channel are i.i.d., we can choose any element to define random vector $\widehat{\textbf{h}}_{k}^{f}\triangleq \widehat{\textbf{h}}_{kr}/ h_{kr}^{b}$ and get the same results, where $ \widehat{\textbf{h}}_{k}^{f}=[\hat{h}_{1k}^{f}, \hat{h}_{2k}^{f}, ..., \hat{h}_{Mk}^{f}]^T$, $\widehat{\textbf{h}}_{kr}=[\hat{h}_{1kr}, \hat{h}_{2kr}, ... ,\hat{h}_{Mkr}]^T$, $r\in\{1, 2, ..., R\}$, and the error  $\textbf{e}_{k}\triangleq \widehat{\textbf{h}}_{k}^{f}-\textbf{h}_{k}^{f}$. 
Then, we can obtain the element of error that follows complex Gaussian distribution with zero mean and variance
	\begin{equation}
	(\sigma_{e,kr}^{LS}(h_{kr}^{b}))^2=\frac{\sigma^2}{|h_{kr}^{b}|^2Dp_{ce}\delta_{k}}\overset{(b)}{=}\frac{K\sigma^2}{|h_{kr}^{b}|^2\alpha p_{ce}\delta_{k}},
	\end{equation}	
	\begin{equation}
	\begin{aligned}
	(\sigma_{e,kr}^{MMSE}(h_{kr}^{b}))^2=\frac{\beta_{k}^2}{|h_{kr}^{b}|^2(1+\frac{\delta_{k}\beta_{k}^2Dp_{ce}}{\sigma^2})}\overset{(c)}{=}\frac{\beta_{k}^2}{|h_{kr}^{b}|^2(1+\frac{\delta_{k}\beta_{k}^2 \alpha p_{ce}}{K\sigma^2})},
	\end{aligned} 
	\end{equation}
where $(b)$ and $(c)$ are hold due to the relation of $\alpha=KD$.

Thus, considering the effect of the backward channel $h_{kr}^{b}$, we have the following Lemma to obtain the conditional distribution of the forward channel $\textbf{h}_{k}^{f}$.	
\begin{lemma}\label{LEMMA1} 
Let $\tilde{\textbf{e}}_{k}=\widehat{\textbf{h}}_{kr}-\textbf{h}_{kr}=\widehat{\textbf{h}}_{kr}-\textbf{G}_{kr}\textbf{h}_{k}^{f}$, where $\textbf{G}_{kr}=diag\{h_{kr}^{b}, h_{kr}^{b},..., h_{kr}^{b}\}\in\mathbb{C}^{M\times M}$. The random vector $\textbf{h}_{k}^{f}=\left[h_{1k}^{f}, h_{2k}^{f}, ... ,h_{Mk}^{f} \right]^T\in\mathbb{C}^{M\times1}\sim\mathcal{CN}\left( \textbf{0}_{M}, \textbf{C}_{\textbf{h}_{k}^{f}}\right)$ and the error vector $\tilde{\textbf{e}}_{k}\sim \mathcal{CN}\left( \textbf{0}_{M}, \textbf{C}_{\tilde{\textbf{e}}_{k}}\right)$. Then, the distribution of the actual forward channel conditioned to the unbiased backscatter channel estimate follows a complex Gaussian distribution, i.e.,
		\begin{equation}
		\begin{aligned}
		\textbf{h}_{k}^{f}|\widehat{\textbf{h}}_{kr},h_{kr}^{b} &\sim \mathcal{CN}\left(\boldsymbol{\mu}_{\textbf{h}_{k}^{f}|\widehat{\textbf{h}}_{kr},h_{kr}^{b}}, \textbf{C}_{\textbf{h}_{k}^{f}|\widehat{\textbf{h}}_{kr},h_{kr}^{b}} \right), 
		\end{aligned}
		\end{equation}
		where the mean vector and covariance matrix are given by
		\begin{eqnarray}
	    &\boldsymbol{\mu}_{\textbf{h}_{k}^{f}|\widehat{\textbf{h}}_{kr},h_{kr}^{b}}=\textbf{C}_{\textbf{h}_{k}^{f}}\textbf{G}_{kr}^{H}\left(\textbf{G}_{kr}\textbf{C}_{\textbf{h}_{k}^{f}}\textbf{G}_{kr}^{H}+ \textbf{C}_{\tilde{\textbf{e}}_{k}}\right)^{-1}\textbf{G}_{kr}\widehat{\textbf{h}}_{k}^{f},\nonumber\\
	    &\textbf{C}_{\textbf{h}_{k}^{f}|\widehat{\textbf{h}}_{kr},h_{kr}^{b}}=\textbf{C}_{\textbf{h}_{k}^{f}}-\textbf{C}_{\textbf{h}_{k}^{f}}\textbf{G}_{kr}^{H}\left(\textbf{G}_{kr}\textbf{C}_{\textbf{h}_{k}^{f}}\textbf{G}_{kr}^{H}+ \textbf{C}_{\tilde{\textbf{e}}_{k}}\right)^{-1}\textbf{G}_{kr}\textbf{C}_{\textbf{h}_{k}^{f}}.\nonumber
	    \end{eqnarray}
\end{lemma}
\begin{IEEEproof}
	Please refer to Appendix~A.
\end{IEEEproof}

Based on the Lemma~\ref{LEMMA1} and the prior distribution of forward channel, the distributions of error vector under LS and MMSE estimation algorithms, i.e., $\textbf{h}_{k}^{f}\sim \mathcal{CN}\left( \textbf{0}_{M}, \beta_{k}\textbf{I}_{M}\right)$, $\tilde{\textbf{e}}_{k}^{LS}\sim \mathcal{CN}\left( \textbf{0}_{M}, (\sigma_{\tilde{e},k}^{LS})^2\textbf{I}_{M}\right)$, $\tilde{\textbf{e}}_{k}^{MMSE}\sim \mathcal{CN}\left(  \textbf{0}_{M}, (\sigma_{\tilde{e},k}^{MMSE})^2\textbf{I}_{M}\right)$, we can obtain the distributions with $h_{kr}^{b}$ as a condition for subsequent analysis
	\begin{equation}
	\begin{aligned}
	\widehat{\textbf{h}}_{kr}|h_{kr}^{b} &\sim \mathcal{CN}\left(\textbf{0}_{M},(|h_{kr}^{b}|^2 \beta_{k}+\sigma_{\tilde {e},k}^2) \textbf{I}_{M}\right), \\
	\widehat{\textbf{h}}_{k}^{f}|h_{kr}^{b} &\sim \mathcal{CN}\left(\textbf{0}_{M},\left( \beta_{k}+\frac{1}{|h_{kr}^{b}|^2}\sigma_{\tilde{e},k}^2\right) \textbf{I}_{M}\right), \\
	\textbf{h}_{k}^{f}|\widehat{\textbf{h}}_{kr},h_{kr}^{b} &\sim \mathcal{CN}\left(\frac{|h_{kr}^{b}|^2\beta_{k}\widehat{\textbf{h}}_{k}^{f}}{|h_{kr}^{b}|^2\beta_{k}+\sigma_{\tilde{e},k}^2}, \frac{\beta_{k}\sigma_{\tilde{e},k}^2}{|h_{kr}^{b}|^2\beta_{k}+\sigma_{\tilde{e},k}^2}\textbf{I}_{M} \right). 
	\end{aligned}
	\end{equation} 

\section{Energy Beamforming Design}\label{sec:design} 
Since the reliable backscatter communication of each hardware-limited tag in multi-user WPBC really depends on the harvested energy and carrier signal from the multi-antenna reader, efficient energy transfer and fair allocation become critical for many self-sustainable IoT applications with numerous backscatter tags~\cite{19Pushing}. To achieve these goals, we utilize energy beamforming and information detection for energy supply and information transfer in WPBC with the estimated BS-CSI. We consider that the reader uses a beamformer with a weighted sum of the normalized estimated BS-CSI and linear detectors (e.g., MRC and ZF receivers). The reasons are as follows: on the one hand, its low computational complexity property is suitable for WPBC system implementation; on the other hand, it is proved to be asymptotically optimal for WET and can adjust the energy allocation weight to balance the rates of all users~[14], [35]. Thus, The beamformer is denoted as  $\boldsymbol{\Phi}=\sum_{k=1}^{K}\sqrt{\zeta_{k}}\frac{\hat{\textbf{h}}_{kr}^*}{\|\hat{\textbf{h}}_{kr}\|}$, where $\zeta_{k}\in{[0,1]}$ $\forall \, k$ such that $\sum_{k=1}^{K}\zeta_{k}=1$.

We note that the energy harvesting rate at the backscatter tag relies on the forward channel while the achievable rate at the reader relies on the backscatter channel. Therefore, there are different optimal energy beams for achievable rate and harvested energy efficiency due to the difference between the forward and backscatter channels. Thus, we formulate the beamforming design as a max-min optimization problem to maximize the minimum rate for all tags by jointly optimizing the energy allocation weights $\boldsymbol{\zeta}=\left[\zeta_{1}, \zeta_{2},..., \zeta_{K} \right]$ of the beamformer $\boldsymbol{\Phi}$, the CE time $\alpha$ and the transmit power of pilot $p_{ce}$, subject to the power consumption constraint as follows.
    \begin{equation}\label{problem formulation}
    \begin{aligned}
    \qquad &  \max\limits_{\boldsymbol{\zeta},\alpha, p_{ce} }\quad \min \limits_{1 \le k \le K} R_{k}(\zeta_{k},\alpha, p_{ce} )\\
    & \begin{array}{r@{\quad}r@{}l@{\quad}l}
    s.t.&&\quad P_{Ek} \geq \rho, \quad \forall k, \\
    &&\quad \sum_{k=1}^{K}\zeta_{k}=1,\\
    &&\quad 0 \leq \alpha \leq T,\\
    &&\quad p_{ce}, \zeta_{k}  \geq 0, \quad \forall k,
    \end{array} 
    \end{aligned} 
    \end{equation}
fg 
where $\rho$ denotes the circuit power consumption rate in backscattering. This problem maximizes the throughput with user-fairness assurance. Due to the non-convexity of the problem, a closed-form analytic solution to~\eqref{problem formulation} does not appear easy to find. Nevertheless, the nonconvex optimization problem can be effectively solved. We observed that the \textit{fminimax} solver in MATLAB multiobjective optimization toolbox can seek a point that minimizes the maximum of a set of multivariable functions using a Sequential Quadratic Programming (SQP) algorithm with modified Hessian and the line search~\cite{algorithm1,algorithm2}. Thus, we can transform our max-min optimization problem to a min-max constraint problem by the identity $ \max\limits_{\boldsymbol{\zeta},\alpha, p_{ce} } \min \limits_{1 \le k \le K} R_{k}= - \min\limits_{\boldsymbol{\zeta},\alpha, p_{ce} } \max \limits_{1 \le k \le K} -R_{k}$ and then use the efficient numerical solver. Note that due to non-convexity of the optimization problem, the global optimum cannot be guaranteed using this numerical solver. Despite that, the simulation results in Section VI demonstrate that the solutions are efficient and verify the performance improvement of the proposed scheme.

\section{Analysis on Energy Harvested Rate and Achievable Rate}\label{sec:analysis}
To optimize beams in a multiuser system as formulated in Eq.~\eqref{problem formulation}, we derive analytical expressions of the energy harvesting rate and the achievable rate by using energy beamforming with the estimated BS-CSI. Additionally, we analyze the effect of the unknown channels on the energy harvesting rate and achievable rate. For analytical tractability, we obtain the bounds on the energy harvesting rate and the achievable rate.

\subsection{Energy Harvesting Rate}
Based on the energy beamforming, the energy harvesting rate~\eqref{energy harvesting rate} can be derived as~\cite{yang2015multi}
    \begin{equation}\label{incident signal power}
    \begin{aligned}
    P_{Ek}&=\eta(1-\delta_{k}) \left( p\zeta_{k}\beta_{k}\mathbb E_{h_{kr}^{b}}\left\lbrace \frac{M\beta_{k}|h_{kr}^{b}|^2+\sigma_{\tilde{e},k}^2}{\beta_{k}|h_{kr}^{b}|^2+\sigma_{\tilde{e},k}^2}\right\rbrace+p\beta_{k}(1-\zeta_{k}) \right)\\
    &=\eta(1-\delta_{k})p\beta_{k}\left[\zeta_{k}\left( M-1\right)\left[ 1-\phi\left(h_{kr}^{b}\right)\right] +1\right],\\
    \end{aligned}
    \end{equation}
where $\phi\left(h_{kr}^{b}\right)=\mathbb E_{h_{kr}^{b}}\left\lbrace1/\left(\beta_{k}|h_{kr}^{b}|^2/\sigma_{\tilde{e},k}^2+1\right)\right\rbrace$.
The result of~\eqref{incident signal power} is obtained by substituting different estimated errors and performing expectation over the unknown backward channel, which takes the fuzziness of $h_{kr}^{b}$ into account. To simplify the calculation, we derive the upper and lower bounds on the incident signal power.

\begin{proposition}\label{boundary}
The boundary of the incident signal power of tag $k$, when energy beamforming is performed using the BS-CSI, is given by
	\begin{equation}
	\begin{aligned}\label{Pbound}
	P_{Ik}^{L}  &\triangleq p\beta_{k} \left[\zeta_{k} (M-1)\mathcal{L}_{1}(\alpha, p_{ce}) +1\right] < P_{Ik}(p_{ce}, p, \alpha, \zeta_{k})\\
	& < p\beta_{k} \left[\zeta_{k} (M-1) \mathcal{L}_{2}(\alpha, p_{ce})+1\right] \triangleq P_{Ik}^{U},
	\end{aligned}
	\end{equation}
	where\\
	\indent$\mathcal{L}_{1}^{LS}(\alpha, p_{ce})=\left[1-\frac{K\sigma^2}{\beta_{k}^2 \alpha p_{ce} \delta_{k}} \ln(1 + \frac{ \beta_{k}^2 \alpha p_{ce} \delta_{k}}{ K\sigma^2 })\right]$,\quad
	\indent$\mathcal{L}_{2}^{LS}(\alpha, p_{ce})=\left[1-\frac{K\sigma^2}{2 \beta_{k}^2 \alpha p_{ce} \delta_{k}} \ln(1 + \frac{ 2 \beta_{k}^2 \alpha p_{ce} \delta_{k}}{ K\sigma^2 })\right]$,\\	\indent$\mathcal{L}_{1}^{MMSE}(\alpha, p_{ce})=\left[1-\frac{K\sigma^2}{K\sigma^2+\beta_{k}^2\alpha p_{ce} \delta_{k}} \ln(1 + \frac{K\sigma^2+\beta_{k}^2\alpha p_{ce} \delta_{k}}{ K\sigma^2 })\right]$,\\
	\indent$\mathcal{L}_{2}^{MMSE}(\alpha, p_{ce})=\left[1-\frac{K\sigma^2}{2(K\sigma^2+\beta_{k}^2\alpha p_{ce} \delta_{k})} \ln(1 + \frac{ 2(K\sigma^2+\beta_{k}^2\alpha p_{ce} \delta_{k}) }{ K\sigma^2 })\right]$.		
\end{proposition}
\begin{IEEEproof}
	Please refer to Appendix~B.
\end{IEEEproof}

Then, we can easily obtain the boundary of the energy harvesting rate according to Eq.~\eqref{energy harvesting rate}. We shall observe numerically in Section VI that the derived upper and lower energy harvesting rate bounds are tight. For analytical tractability, we will use these tight boundaries to simplify the operation.

\subsection{Ergodic Achievable Rate}
We consider two conventional linear detectors MRC and ZF. Unlike traditional MU-MIMO systems, we cannot directly estimate the backward channel for information detection. Thus we use backscatter channel instead of backward channel, i.e.,
    \begin{align}\label{receiver}
    \textbf{Q}= \begin{cases}\widehat{\textbf{H}}_{m}, & for\quad MRC,\\\widehat{\textbf{H}}_{m} \left(\widehat{\textbf{H}}_{m}^{H}\widehat{\textbf{H}}_{m}\right)^{-1}, & for\quad ZF,\end{cases}
    \end{align}
where $\widehat{\textbf{H}}_{m}=[\widehat{\textbf{h}}_{m1},\widehat{\textbf{h}}_{m2}, ..., \widehat{\textbf{h}}_{mK}]$.

To analyze the achievable rate, we consider F-CSI as an unknown parameter to obtain the desired B-CSI through BS-CSI. Mathematically, we define a random vector $\widehat{\textbf{h}}_{k}^{b}\triangleq \widehat{\textbf{h}}_{mk}/ h_{mk}^{f}$, where $\widehat{\textbf{h}}_{k}^{b}=[\hat{h}_{k1}^{b}, \hat{h}_{k2}^{b}, ..., \hat{h}_{kR}^{b}]^T$,  $\widehat{\textbf{h}}_{mk}=[\hat{h}_{mk1}, \hat{h}_{mk2}, ... ,\hat{h}_{mkR}]^T$, $m\in\{1, 2, ..., M\}$, and the error  $\boldsymbol{\varepsilon}\triangleq \widehat{\textbf{H}}^{b}-\textbf{H}^{b}$. Furthermore,
 we can obtain the distributions of $\widehat{\textbf{h}}_{mk}|h_{mk}^{f}$, $\widehat{\textbf{h}}_{k}^{b}|h_{mk}^{f}$, $\textbf{h}_{k}^{b}|\widehat{\textbf{h}}_{mk},h_{mk}^{f}$ according to the Lemma~\ref{LEMMA1} and the prior distribution of backward channel, the analytical results of channel estimation errors under LS and MMSE estimation algorithms in Section~\ref{sec:estimator}.
\begin{equation}
\begin{aligned}
\widehat{\textbf{h}}_{mk}|h_{mk}^{f} &\sim \mathcal{CN}\left(\textbf{0}_{R},(|h_{mk}^{f}|^2 \beta_{k}+\sigma_{\tilde {e},k}^2) \textbf{I}_{R}\right), \\
\widehat{\textbf{h}}_{k}^{b}|h_{mk}^{f} &\sim \mathcal{CN}\left(\textbf{0}_{R},\left( \beta_{k}+\frac{1}{|h_{mk}^{f}|^2}\sigma_{\tilde{e},k}^2\right) \textbf{I}_{R}\right), \\
\textbf{h}_{k}^{b}|\widehat{\textbf{h}}_{mk},h_{mk}^{f} &\sim \mathcal{CN}\left(\frac{|h_{mk}^{f}|^2\beta_{k}\widehat{\textbf{h}}_{k}^{b}}{|h_{mk}^{f}|^2\beta_{k}+\sigma_{\tilde{e},k}^2}, \frac{\beta_{k}\sigma_{\tilde{e},k}^2}{|h_{mk}^{f}|^2\beta_{k}+\sigma_{\tilde{e},k}^2}\textbf{I}_{R} \right). 
\end{aligned}
\end{equation} 

With imperfect CSI, the received signal of tag $k$ after using a linear detector is rewritten as
    \begin{equation}
    r_{k} = \textbf{q}_{k}^{H} \widehat{\textbf{h}}_{k}^{b} x_{k} + \sum_{i=1,i\neq k}^{K} \textbf{q}_{k}^{H} \widehat{\textbf{h}}_{i}^{b} x_{i} - \sum_{i=1}^{K} \textbf{q}_{k}^{H} \boldsymbol{\varepsilon}_{i} x_{i}+ \textbf{q}_{k}^{H} \boldsymbol{\mu},
    \end{equation}
where $\widehat{\textbf{h}}_{k}^{b}$ and $\boldsymbol{\varepsilon}_{k}$ are the $k$-th column of $\widehat{\textbf{H}}^{b}$ and $\boldsymbol{\varepsilon}$, respectively. 
Furthermore, the ergodic achievable rate of the information transmission from tag $k$ is given by
    \begin{equation}\label{rate}
    R_{k} \triangleq \mathbb{E} \left\lbrace \log_{2}\left( 1 +  \gamma_{k} \right) \right\rbrace,  
    \end{equation}
where the signal-to-interference-plus-noise-ratio(SINR) is expressed as
    \begin{equation}
    \begin{aligned}
    \gamma_{k} & =   \frac{ p_{k} |\textbf{q}_{k}^{H} \widehat{\textbf{h}}_{k}^{b}|^2 }{\sum_{i=1,i\neq k}^{K}  p_{i} |\textbf{q}_{k}^{H} \widehat{\textbf{h}}_{i}^{b}|^2  + |\textbf{q}_{k}^{H} \textbf{q}_{k}|\sum_{i=1}^{K}  p_{i}  \sigma_{\epsilon,mi}^2(h_{mi}^{f})+ |\textbf{q}_{k}^{H} \textbf{q}_{k}|\sigma^2}.
    \end{aligned} 
    \end{equation}

The exact expression for the rate $R_{k}$ in~\eqref{rate} is analytically intractable. Since $f(x)=\log(1+ \frac{1}{x})$ is a convex function, by utilizing Jensen's inequality, the lower bound on the ergodic achievable rate is given by
	\begin{equation}\label{lower boound}
	\begin{aligned}
	R_{k} \ge \tilde{R}_{k} &\triangleq \log_{2}(1 + \tilde{\gamma}_{k})
	\\&= \log_{2}\left(1 + \left(\mathbb{E} \left \lbrace \frac{\sum_{i=1,i\neq k}^{K}  p_{i} |\textbf{q}_{k}^{H} \widehat{\textbf{h}}_{i}^{b}|^2  + |\textbf{q}_{k}^{H} \textbf{q}_{k}|\sum_{i=1}^{K}  p_{i}  \sigma_{\epsilon,mi}^2(h_{mi}^{f})+ |\textbf{q}_{k}^{H} \textbf{q}_{k}|\sigma^2}  { p_{k} |\textbf{q}_{k}^{H} \widehat{\textbf{h}}_{k}^{b}|^2 }\right \rbrace \right)^{-1}\right).
	\end{aligned} 
	\end{equation} 
The result of~\eqref{lower boound} is obtained by performing expectation over the unknown backward and forward channels, which takes the effect of unknown channels into account.

\underline{MRC Receiver.}
For MRC receiver, according to~\eqref{receiver}, we have $\textbf{q}_{k} = \widehat{\textbf{h}}_{mk}$. We can obtain the achievable rate in the following Proposition. 
\begin{proposition}\label{mrc_rate}
For MRC detection and $R \ge 2$, the general expression of achievable rate under two different channel estimation methods is given by 
	\begin{equation}
	\begin{aligned}
	\tilde{R}^{MRC}_{k} &= \log_{2}(1 + \tilde{\gamma}^{MRC}_{k}),\\
	\end{aligned} 
	\end{equation} 
	where $p_{k} = \delta_{k} P_{Ik}$, and the SINR
		\begin{equation}\label{SINR}
		\begin{aligned}
		\vspace{-0.5cm}
		\tilde{\gamma}^{MRC}_{k} = \frac{p_{k} (R-1)\beta_{k}}{\left(1-\phi\left(h_{mk}^{f}\right)\right) \bigg( \sum_{i=1,i\neq k}^{K} p_{i}\beta_{i}  + 2\sum_{i=1,i\neq k}^{K} p_{i} \sigma_{\tilde{e},i}^2 \frac{\Gamma(0,\frac{\tau}{\beta_{i}})}{\beta_{i}} + \sigma^2 \bigg)  + p_{k}\beta_{k} \phi\left(h_{mk}^{f}\right)}.
		\end{aligned} 
		\end{equation}
\vspace{-1cm}
\end{proposition}
\begin{IEEEproof}
	Please refer to Appendix~C.
\end{IEEEproof}

\underline{ZF Receiver.}
For ZF receiver, according to~\eqref{receiver}, we have $\textbf{Q} = \widehat{\textbf{H}}_{m} \left(\widehat{\textbf{H}}_{m}^{H}\widehat{\textbf{H}}_{m}\right)^{-1}$. We can obtain the achievable rate in the following Proposition.  
\begin{proposition}\label{zf_rate}
With ZF detection and $R \ge K+1$, we can obtain the general expression of achievable rate under two different channel estimation methods as
	\begin{equation}
	\begin{aligned}
	\tilde{R}^{ZF}_{k} &= \log_{2}(1 + \tilde{\gamma}^{ZF}_{k}),\\
	\end{aligned} 
	\end{equation} 
	where the SINR 
		\begin{equation}\label{SINR_ZF}
		\begin{aligned}
		\vspace{-0.5cm}
		\tilde{\gamma}^{ZF}_{k} = \frac{p_{k} (R-K)\beta_{k}}{\left( 1 - \phi\left(h_{mk}^{f}\right)\right) \left(  \sum_{i=1,i\neq k}^{K}  p_{i} \frac{\sigma_{\tilde{e},i}^2}{\beta_{i}}\Gamma(0, \frac{\tau}{\beta_{i}}) + \sigma^2 \right)+ p_{k}\beta_{k} \phi\left(h_{mk}^{f}\right)}.
		\end{aligned} 
		\end{equation} 
\vspace{-1cm}
\end{proposition}
\begin{IEEEproof}
	Please refer to Appendix~D.
\end{IEEEproof}

We also note that the SINR appears analytically intractable. For analytical tractability, we obtain the lower bound as 
	\begin{equation}\label{lower SINR}
	\begin{aligned}
	&\tilde{\gamma}^{MRC}_{kL} = \frac{\delta_{k} P_{Ik}^{L} (R-1)\beta_{k}}{\mathcal{L}_{2}(\alpha, p_{ce}) \left( \sum_{i=1,i\neq k}^{K}  \delta_{i} P_{Ii}^{U} \beta_{i} + 2\sum_{i=1,i\neq k}^{K} \delta_{i} P_{Ii}^{U} \frac{\sigma_{\tilde{e},i}^2}{\beta_{i}} e^{-\frac{\tau}{\beta_{i}}} \ln(1+\frac{\beta_{i}}{\tau}) + \sigma^2 \right) + \delta_{k} P_{Ik}^{U}\frac{\sigma_{\tilde{e},k}^2}{\beta_{k}}\ln(1+\frac{\beta_{k}^2}{\sigma_{\tilde{e},k}^2})}\\
	&\tilde{\gamma}^{ZF}_{kL} = \frac{\delta_{k} P_{Ik}^{L} (R-K)\beta_{k}}{\mathcal{L}_{2}(\alpha, p_{ce}) \left( \sum_{i=1,i\neq k}^{K} \delta_{i} P_{Ii}^{U} \frac{\sigma_{\tilde{e},i}^2}{\beta_{i}} e^{-\frac{\tau}{\beta_{i}}} \ln(1+\frac{\beta_{i}}{\tau}) + \sigma^2 \right) + \delta_{k} P_{Ik}^{U}\frac{\sigma_{\tilde{e},k}^2}{\beta_{k}}\ln(1+\frac{\beta_{k}^2}{\sigma_{\tilde{e},k}^2})}.
	\end{aligned}
	\end{equation}
We shall observe numerically in Section VI that the analytically-obtained rate bounds approach that obtained from simulation, especially for ZF. Thus, the lower bounds $\tilde{R}_{kL}$ are tight for WPBC systems. In the sequel, we take the lower bound $\tilde{R}_{kL}$ as the achievable rate, and $P_{Ek}^{L}$ as instantaneous energy harvesting rate in Eq.~\eqref{problem formulation} to complete the beamforming design.

\subsection{Beamforming with Perfect CSI and Omnidirectional Transmission}
To verify the performance gain of the proposed scheme, we give two benchmarks, i.e., the case of beamforming with perfect CSI and the case of omnidirectional transmission. 

\underline{Beamforming with Perfect CSI.} 
In this case, we assume the reader has perfect knowledge of F-CSI for energy beamforming and B-CSI for information detection. Thus, the step of the backscatter CE can be removed. With the energy beamforming, the incident signal power of tag $k$ can be derived as 
    \begin{equation}\label{incident signal power_P-CSI}
    \begin{aligned}
    P_{Ik}^{P}&=p\beta_{k} \left[M\zeta_{k}+(1-\zeta_{k})\right].
    \end{aligned}
    \end{equation}
Then, the instantaneous energy harvesting rate is $P_{Ek}^{P}=\eta(1-\delta_{k})P_{Ik}^{P}$. Moreover, the achievable rate is given by
    \begin{align}
    \tilde{R}^{P}_{k}= \begin{cases}\log_{2}\left(1 +\frac{p\delta_{k}\beta_{k}^2(R-1)\left[ M\zeta_{k}+(1-\zeta_{k})\right]}{\sum_{i=1,i\neq k}^{K}p\delta_{i}\beta_{i}^2\left[ M\zeta_{i}+(1-\zeta_{i})\right]+ \sigma^2}\right), & for\quad MRC,\\\log_{2}\left(1 +\frac{p\delta_{k}\beta_{k}^2(R-K)\left[ M\zeta_{k}+(1-\zeta_{k})\right]}{\sigma^2}\right), & for\quad ZF.\end{cases}
    \end{align}

\underline{Omnidirectional Transmission.} 
In this case, the reader performs omnidirectional transmission without energy beamforming. Channel estimation is only used for information detection. We can obtain the incident signal power of tag $k$
    \begin{equation}\label{incident signal power_O}
    P_{Ik}^{O}=p\beta_{k} \left[ 1-\frac{\left( M-1\right)}{M}\mathbb E_{h_{kr}^{b}}\left\lbrace\frac{1}{\frac{\beta_{k}|h_{kr}^{b}|^2}{\sigma_{\tilde{e},k}^2}+1}\right\rbrace \right].
    \end{equation}
Then, by changing $P_{Ik}$ to $P_{Ik}^{O}$ in Propositions 2 and 3, we can obtain the achievable rate as
    \begin{align}
    \tilde{R}^{O}_{k}= \begin{cases}\log_{2}\left(1 + \frac{p_{k}^{O} (R-1)\beta_{k}} {\left(1-\phi\left(h_{mk}^{f}\right)\right) \bigg( \sum_{i=1,i\neq k}^{K} p_{i}^{O}\beta_{i}  + 2\sum_{i=1,i\neq k}^{K} p_{i}^{O} \sigma_{\tilde{e},i}^2 \frac{\Gamma(0,\frac{\tau}{\beta_{i}})}{\beta_{i}} + \sigma^2 \bigg)  + p_{k}^{O}\beta_{k} \phi\left(h_{mk}^{f}\right)}\right), & for\quad MRC,\\\log_{2}\left(1 + \frac{p_{k}^{O} (R-K)\beta_{k}} {\left( 1 - \phi\left(h_{mk}^{f}\right)\right) \left(  \sum_{i=1,i\neq k}^{K}  p_{i}^{O} \frac{\sigma_{\tilde{e},i}^2}{\beta_{i}}\Gamma(0, \frac{\tau}{\beta_{i}}) + \sigma^2 \right)+ p_{k}^{O}\beta_{k} \phi\left(h_{mk}^{f}\right)}\right), & for\quad ZF,\end{cases}
    \end{align}
where $p_{k}^{O} = \delta_{k} P_{Ik}^{O}$.       

\section{Numerical Results}\label{sec:evaluation}

In this section, we conduct simulations to quantify the efficacy of the proposed scheme. Unless explicitly stated, the default system parameter values are $M=R=4$. The average transmit power of the reader is set to be 2 W. We consider two backscatter tags and assume the duration of each time block is $T = 200$ symbol periods. We set the noise power as $- 100$ dBm. We also use the long-term fading model $\beta_{k} = \frac{A_e}{4\pi d_{k}^2}$~\cite{yang2015multi}, where the distance $d_{1} = 4$ m, $d_{2} =6$ m, $d_{3} =8$ m, $d_{4} =6$ m, $d_{5} =5$ m, and $d_{6} =10$ m. The denominator of the fading model is the effective aperture $A_e$ of the tag antenna, which is defined as the constant of proportionality of the power collected to the density of power impinging on the tag. The effective aperture for an isotropic antenna is given by $A_e=\frac{(3\times 10^8)^2}{4\pi f^2}$ m$^2$~\cite{07UHFRFID} and the frequency is 915 MHz. The energy conversion coefficient is $\eta = 0.65$ and the power reflection coefficient of two tags is $\delta_{1} = \delta_{2} = 0.25$, which implies that about $25\%$ incident power is backscattered~\cite{gang2018Modulation}. We consider the circuit power consumption rates as $\rho = 8.9~\mu$W~\cite{lu2018wireless}.

\begin{figure}
	\centering
	\begin{minipage}[b]{0.45\textwidth}\centering
		\includegraphics[width=0.99\textwidth]{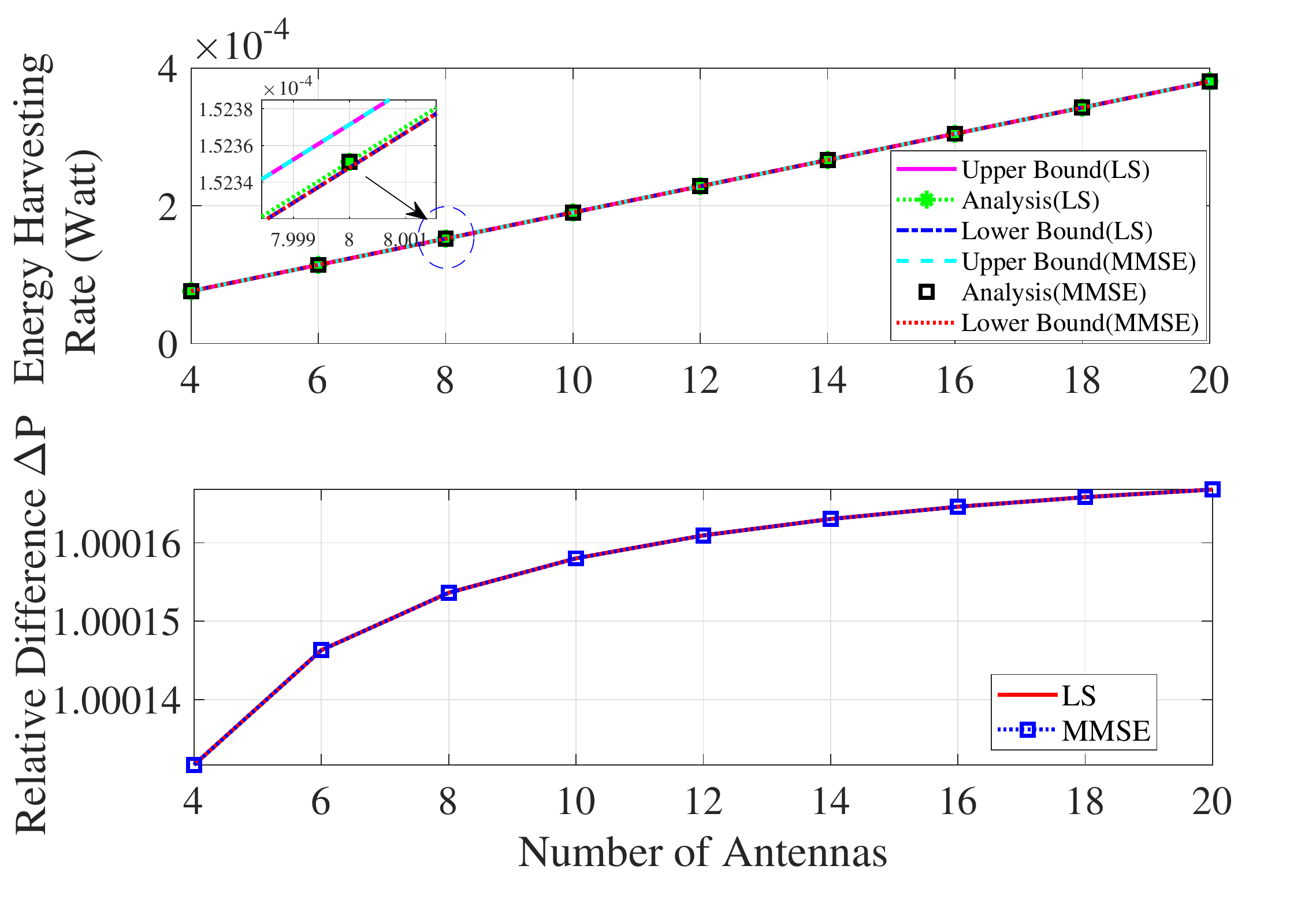}\vspace{-0.3cm}
		\caption{Energy harvesting rate and its relative difference vs. $M$ for simulation and bounds.}
		\label{fig:figurePE}
	\end{minipage}
	\begin{minipage}[b]{0.45\textwidth}\centering
		\includegraphics[width=0.99\textwidth]{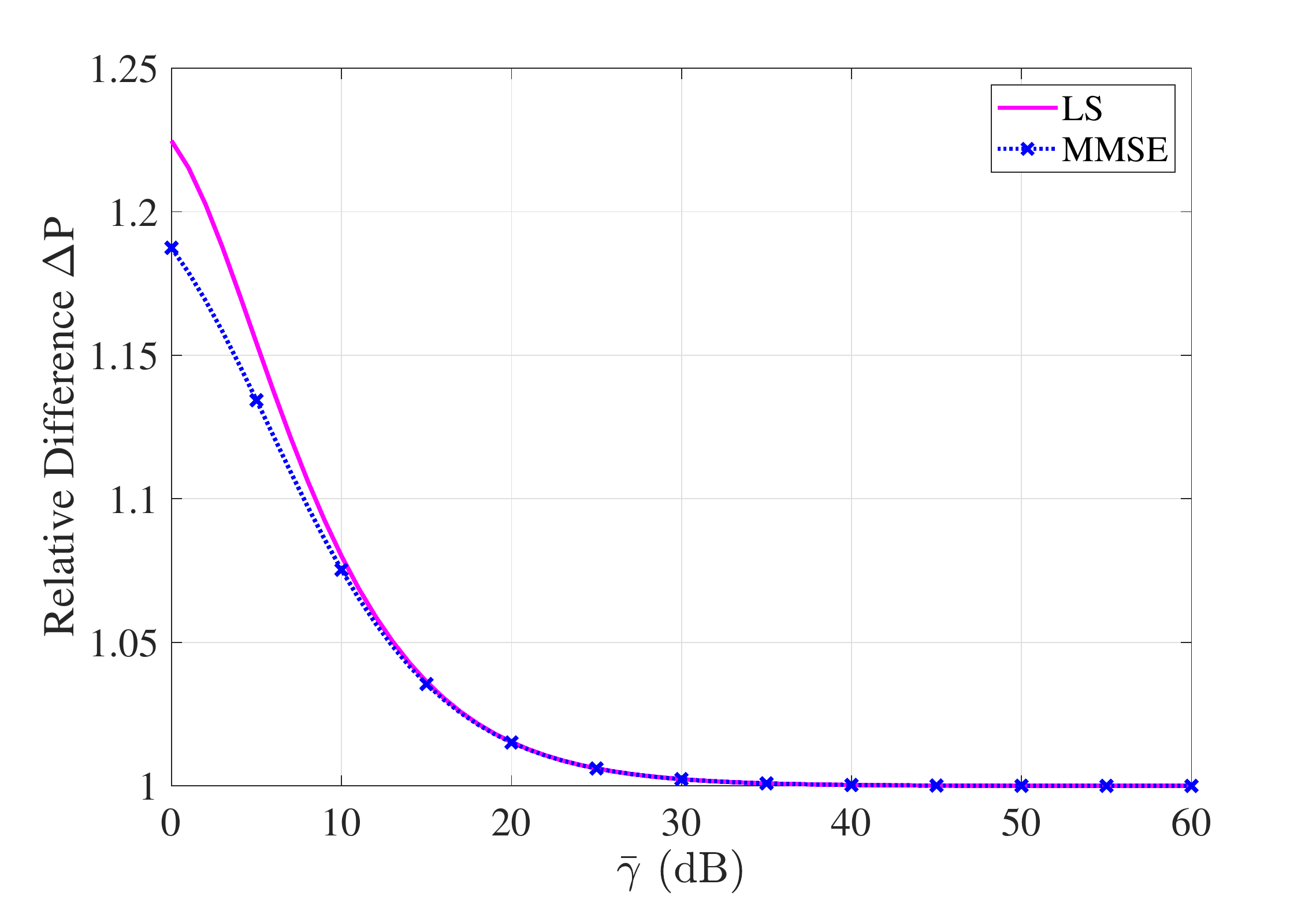}\vspace{-0.3cm}
		\caption{Relative difference $\Delta P$ vs. SNR $\bar{\gamma} \triangleq \bar{\beta}^2 \alpha p_{ce} \delta/K \sigma^2$.}
		\label{fig:figurePER}
	\end{minipage}
\end{figure}

\begin{figure}
	\centering
	\begin{minipage}[b]{0.45\textwidth}\centering
		\includegraphics[width=0.99\textwidth]{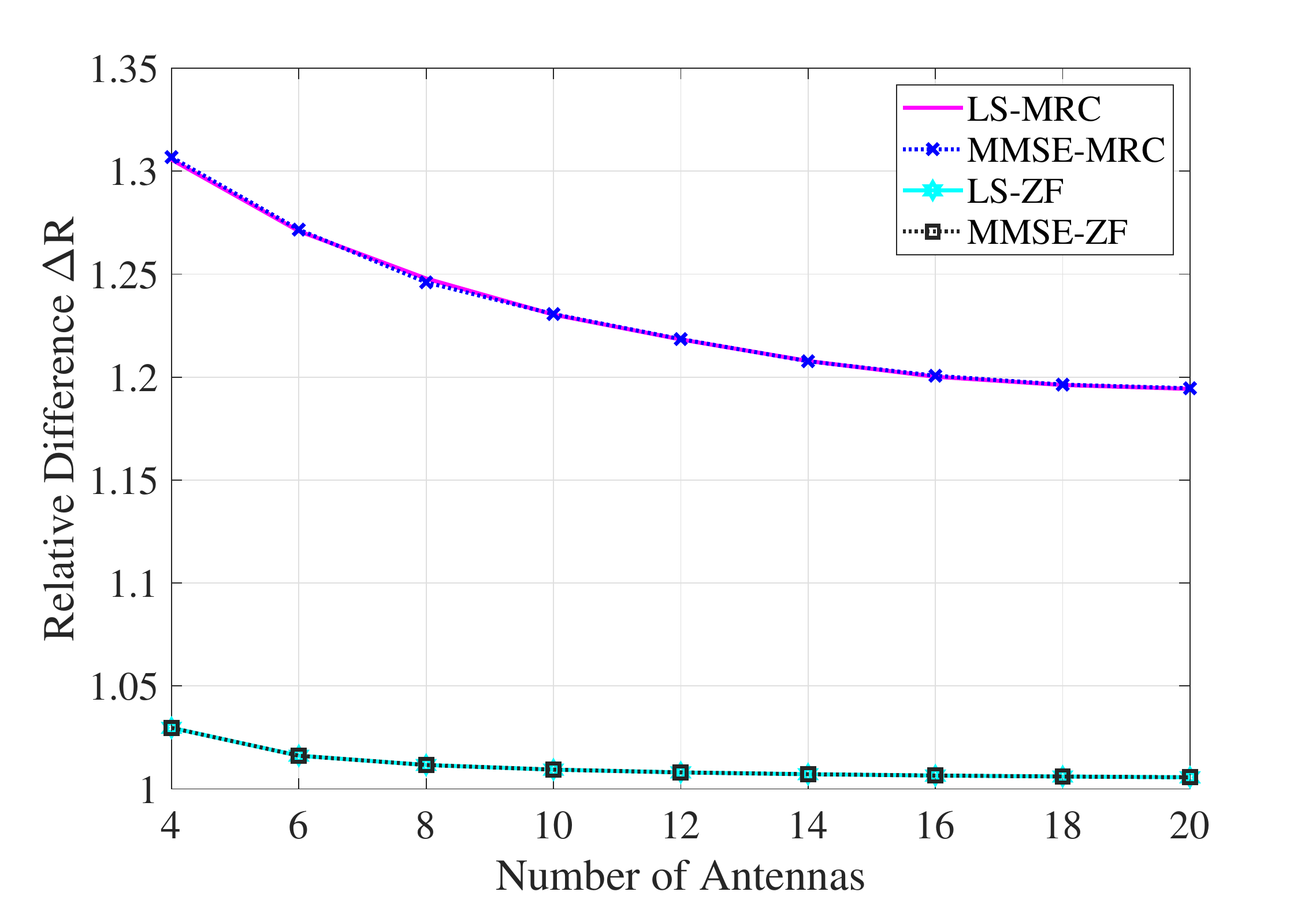}\vspace{-0.3cm}
		\caption{Relative difference $\Delta R$ vs. the receive antennas for simulation and bounds.}
		\label{fig:figureR}
	\end{minipage}
	\begin{minipage}[b]{0.45\textwidth}\centering
		\includegraphics[width=0.99\textwidth]{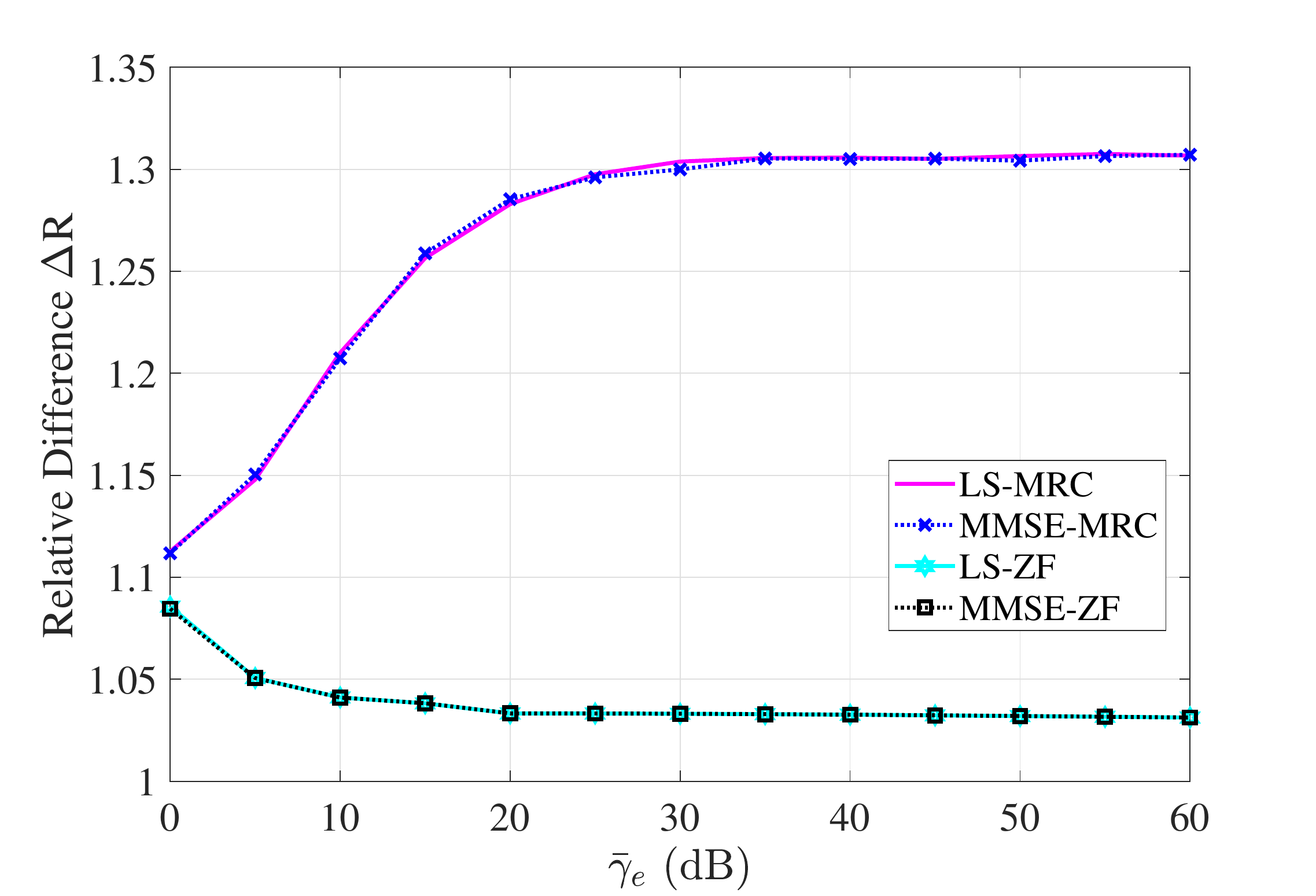}\vspace{-0.3cm}
		\caption{Relative difference $\Delta R$ vs. the average effective SNR $\bar{\gamma}_e \triangleq \bar{\beta}^2 / \sigma^2 $.}
		\label{fig:figureRR}
	\end{minipage}
\end{figure}
We first conduct experiments to validate the tightness of our proposed energy harvesting rate bounds and achievable rate bounds. Fig.~\ref{fig:figurePE} shows both the relative difference and the absolute difference between the simulated energy harvesting rate and the proposed analytical bounds for one tag with LS and MMSE estimators. The gaps between the upper and lower bounds with LS and MMSE estimators are less than $0.017\%$ aginst the varying number of antennas. Moreover, we verify the tightness of the energy harvesting rate bounds via Fig.~\ref{fig:figurePER}, which plots the relative difference $\Delta P \triangleq P_{Ik}^U/P_{Ik}^L=P_{Ek}^U/P_{Ek}^L$ referring to Eq.~\eqref{Pbound} in Proposition~\ref{boundary}, aginst increasing SNR $\bar{\gamma} \triangleq \bar{\beta}^2 \alpha p_{ce} \delta/K \sigma^2$, where $\bar{\beta} \triangleq \frac{1}{K}\sum_{k=1}^{K} \beta_{k}$, with $K=1$. As observed from Fig. 3, the gap is quite sensitive to the SNR. However, the result validates the quality of the energy harvesting rate bounds with an acceptable average gap between the lower and upper bounds of less than $0.6\%$ and $7.5\%$ for $\bar{\gamma}>25$ dB and $\bar{\gamma}>10$ dB, respectively. Only when $\bar{\gamma}$ approaches 0, the gap is around $20\%$ large. Therefore, the derived upper and lower energy harvesting rate bounds are very tight when SNR is relatively high.

\begin{figure}
	\centering
	\begin{minipage}[b]{0.45\textwidth}\centering
		\includegraphics[width=0.99\textwidth]{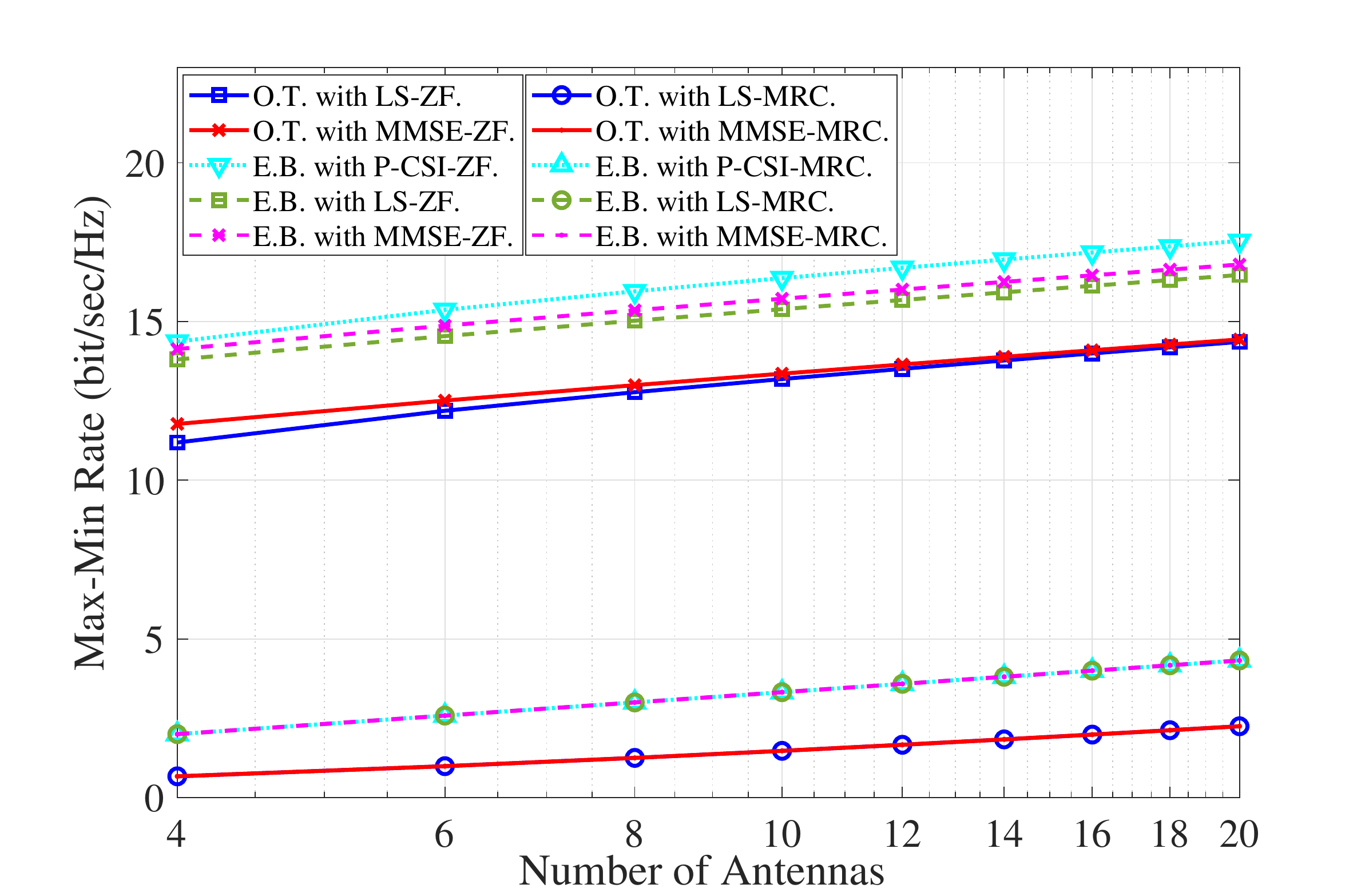}\vspace{-0.3cm}
		\caption{Max-min rate vs. $R$ for three schemes.}
		\label{fig:figure2}
	\end{minipage}
	\begin{minipage}[b]{0.45\textwidth}\centering
		\includegraphics[width=0.99\textwidth]{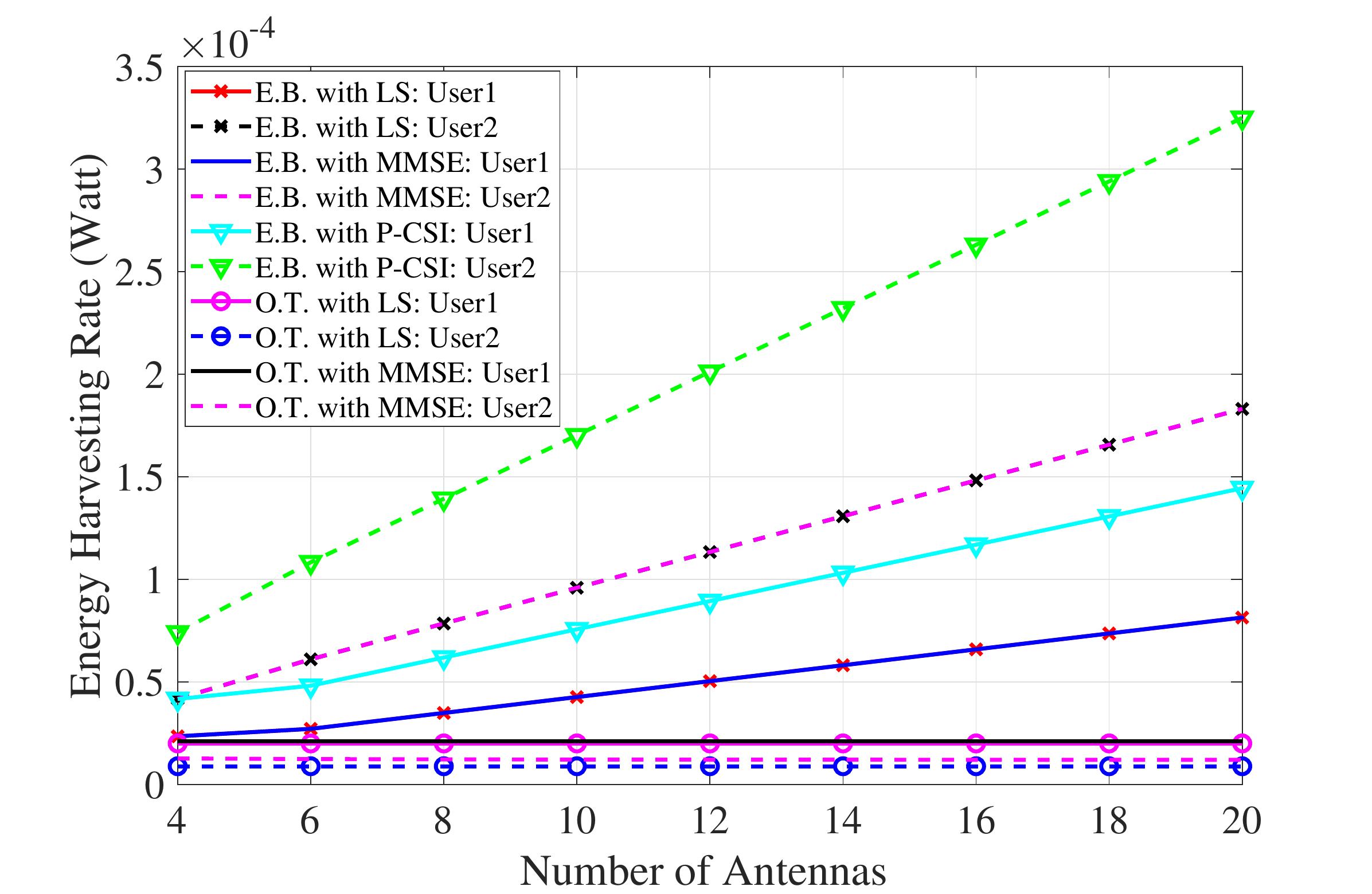}\vspace{-0.3cm}
		\caption{Energy harvesting rate vs. $M$ for three schemes.}
		\label{fig:figure3}
	\end{minipage}
\end{figure}

After that we focus on verifying the tightness of the derived closed-form lower bound $\tilde{R}$ for the achievable rate during the information transfer phase. We plot the relative difference $\Delta R \triangleq R_{sim}/\tilde{R} $ for the achievable rate under the varying number of antennas for the MRC and ZF receivers. From Fig. 4, we notice that the gap between the analytically-obtained rate bounds and the simulation results for MRC is sensitive to $R$. MRC yields an average gap of $23\%$. However, the gap for ZF is no more than $3\%$. Although the gap of MRC is large at small $R$, the match between the analytically-obtained rate bounds and simulation results is tight for ZF receiver even though the number of antennas is small. Moreover, we plot the relative errors $\Delta R$ against increasing average effective SNR $\bar{\gamma}_e \triangleq \bar{\beta}^2 / \sigma^2 $ in Fig. 5. We observe that the gap of MRC is more sensitive to $\bar{\gamma}_e$ than that of ZF. The gap for the MRC case degrades to $11\%$ as $\bar{\gamma}_e$ approaches 0 dB and is bounded by $30\%$ for large $\bar{\gamma}_e$. Besides, the result validates the quality of the achievable rate bounds with an acceptable average gap between the lower bound and simulation result of less than $5\%$ for $\bar{\gamma}_e>5$ dB for ZF. Thus, the match for ZF case is tight. Therefore, in the following, we will use these bounds for all numerical work.

We next compare the proposed scheme with the two benchmarks to illustrate its efficiency.  We have considered the max-min rate as a validation metric for evaluating the goodness of three transmission schemes. Fig.~\ref{fig:figure2} shows the max-min achievable rate versus the different numbers of reader's receive antennas ranging from 4 to 20, while the number of transmit antennas is 4. In Fig.~\ref{fig:figure2}, we can observe that the max-min rate of the proposed scheme is significantly higher than that of omnidirectional transmission, but slightly lower than that obtained by energy beamforming with perfect CSI. Specifically, the achievable rate of proposed scheme approaches more than $90\%$ of the rate limit achieved by the ideal case with perfect CSI for both receivers. Moreover, we see that the max-min rate is increased significantly by $200\%$ compared to omnidirectional transmissions for the MRC receiver, when $M=R=4$.
	
\begin{figure}
	\centering  
	\subfigure[\scriptsize MRC Receiver]
	{\label{fig:figure41}\includegraphics[width=0.45\textwidth]{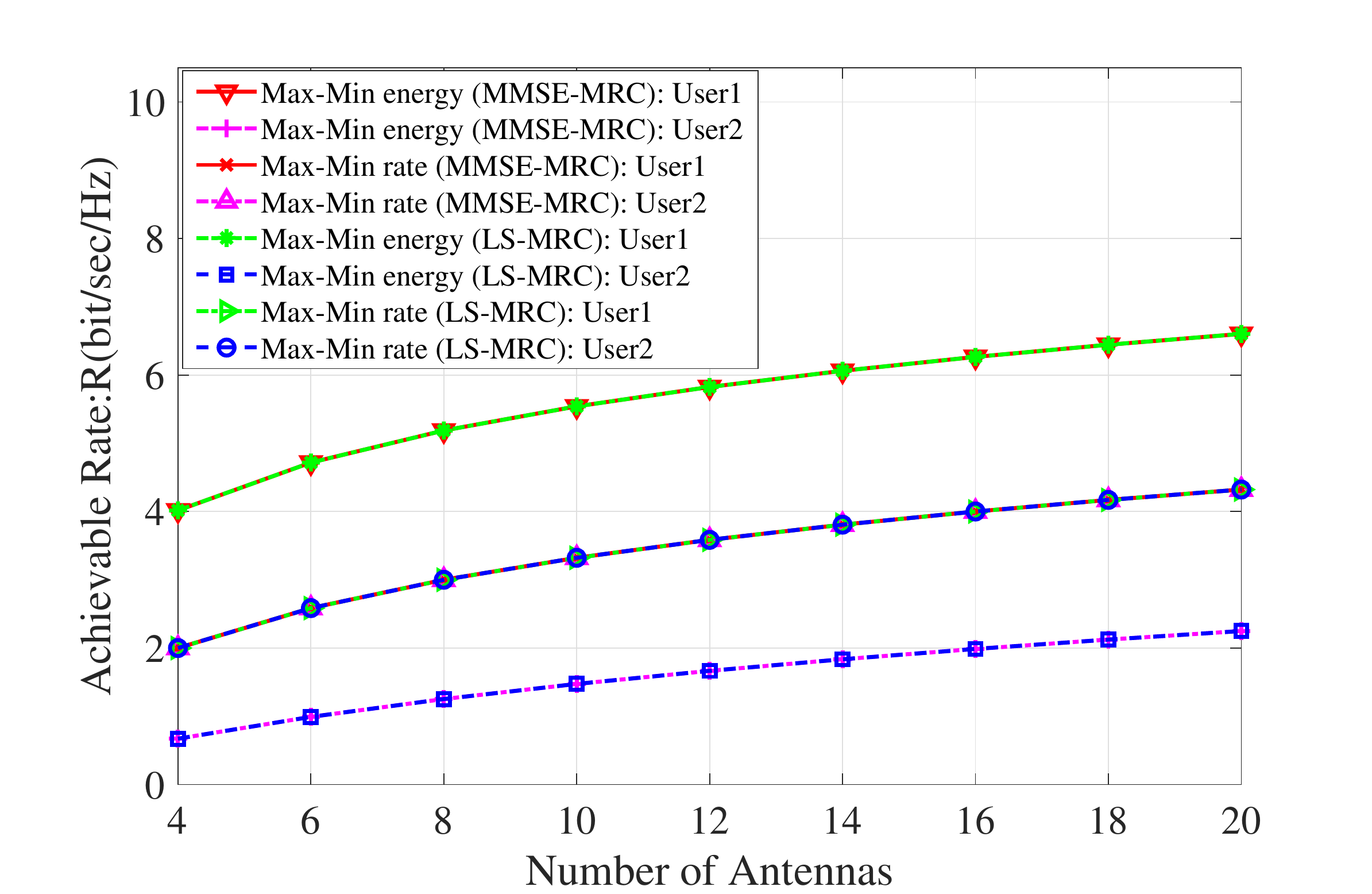}}
	\subfigure[\scriptsize ZF Receiver]
	{\label{fig:figure51} \includegraphics[width=0.45\textwidth]{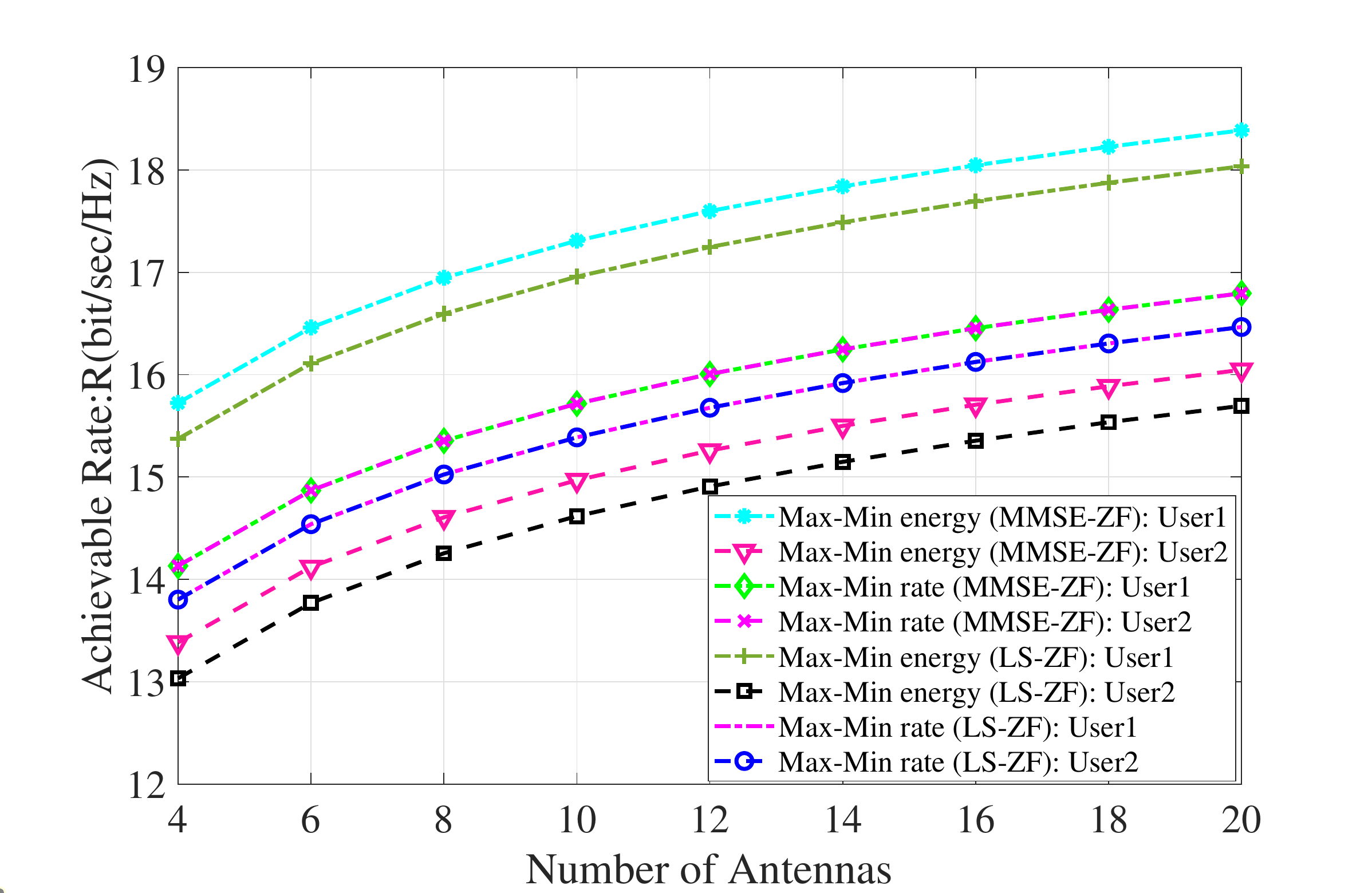}}
	\caption{Achievable rate vs. $R$ for different optimization goals.}
	\label{fig:figure45}
	\vspace{-0.3cm}
\end{figure}
	
Besides, when comparing MRC and ZF receivers in Fig.~\ref{fig:figure2}, we observe that the max-min achievable rate for the ZF receiver is higher than that for the MRC receiver. In the case of using ZF receiver, there is a gap between the performance of LS and that of MMSE. In the case of using MRC receiver, LS estimator achieves comparable performance compared with MMSE estimator. Since the achievable rate is affected by the inter-user interference and estimation errors. However, the ZF receiver has a good suppression effect on the inter-user interference. Thus, the relative effect of estimation errors as compared to the inter-user interference is more prominent when using the ZF receiver. This coincides with the analytic result in Propositions 2 and 3.

Fig.~\ref{fig:figure3} plots the energy harvesting rate of different schemes with LS and MMSE estimators versus the different numbers of reader's transmit antennas ranging from 4 to 20, while the number of receive antennas is 4. We consider the MRC receiver since the selection of the uplink information detection scheme has little influence on the downlink energy transfer. From Fig.~\ref{fig:figure3}, we also observe that the energy harvesting rate of the proposed scheme has a large gain compared to that of omnidirectional transmission. Furthermore, as $M$ increases, the energy harvesting rate by using energy beamforming increases, which corroborates the analysis in Section V.A, while the energy harvesting rate by omnidirectional transmission remains to be a small constant, which coincides with the analytic result in Section V.C. Due to the extremely low energy harvesting rate of omnidirectional transmission, it is difficult to activate the backscatter tag for communication when the transmission power at the reader is low. However, beamforming can greatly increase the energy harvesting rate to activate tags and ensure uninterrupted backscattering of tags. Based on the above observations in Figs.~\ref{fig:figure2} and~\ref{fig:figure3}, our system enables efficient energy and information transfer by exploiting energy beamforming.

Then, we compare our energy beamforming design and the conventional max-min energy-based design for different numbers of antennas deployed at the reader. Note that our scheme maximizes the minimum rate for all backscatter tags subject to the power consumption constraint. Fig.~\ref{fig:figure45} indicates that the minimum achievable rate of the proposed energy beamforming design is significantly higher than that of previous max-min energy-based design for both two different receivers. These results show that our energy beamforming scheme can effectively guarantee the reliable communications and fairness of all tags. Additionally, we can observe that the gap between the performance of LS and that of MMSE in Fig.~\ref{fig:figure51} is larger compared with Fig.~\ref{fig:figure41}. This is because the relative effect of estimation errors as compared to the inter-user interference is more prominent when using the ZF receiver.
\begin{figure}
	\centering
	\includegraphics[width=0.5\linewidth]{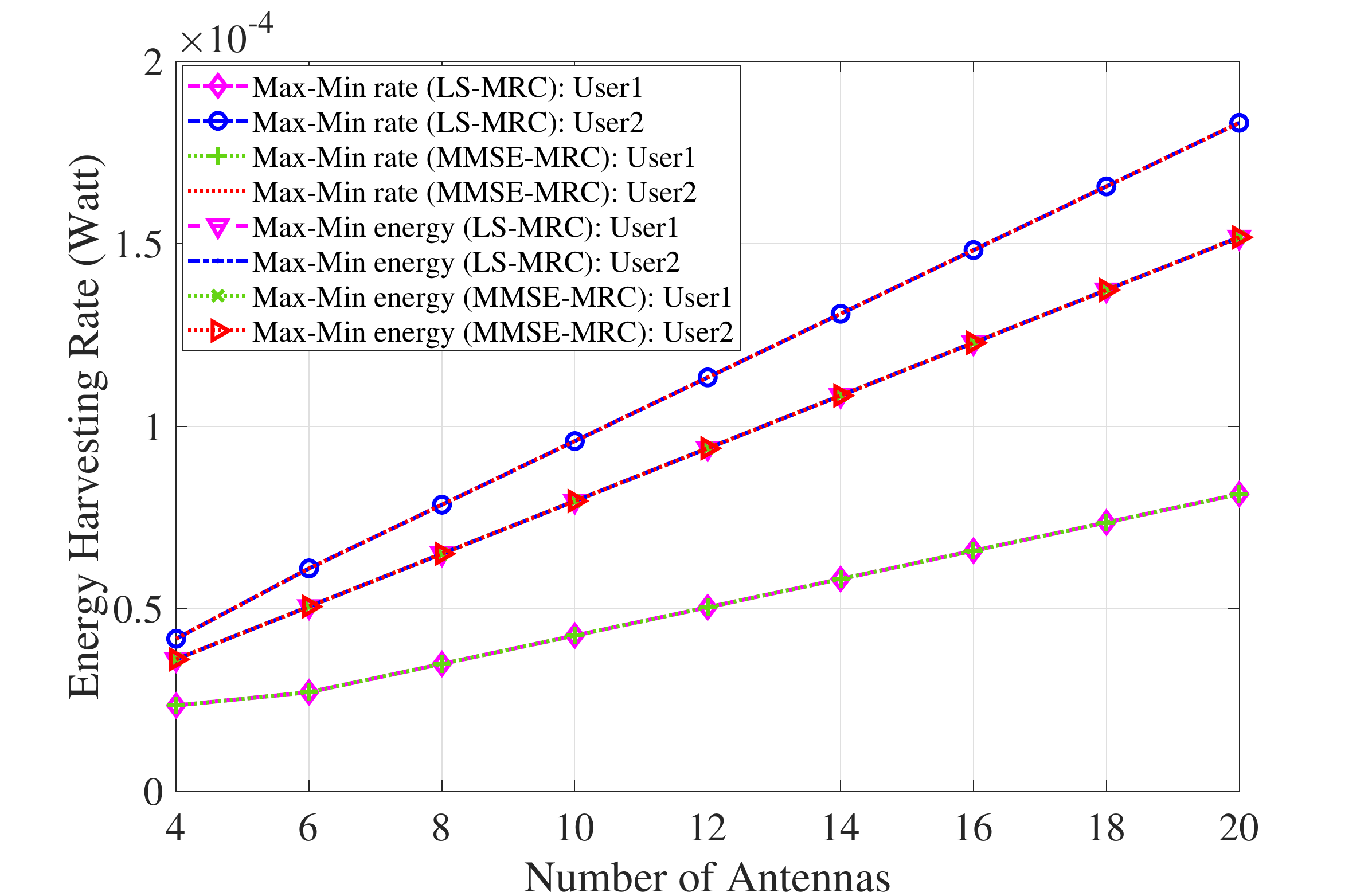}
	\caption{Energy harvesting rate vs. $M$ for different optimization goals.}
	\label{fig:figure6}
\end{figure}

Fig.~\ref{fig:figure6} depicts the energy harvest rate of above two optimal energy beamforming  designs. For our energy beamforming design, even though the energy harvesting rate of tag $1$ is low, it can also reach the same achievable rate as tag $2$, as shown in Fig.~\ref{fig:figure41}. This is because our energy beamforming design suppresses the double near-far effect~\cite{yang2015throughput} by adjusting the beams to deliver more energy to the farther user. The beamforming design which maximizes the minimum energy can ensure the energy harvesting rates fairness. However, due to the double near-far effect, the rate of tag $2$ is $83\%$ lower than that of tag $1$, when $M=R=4$. Thus, it is wasteful for the portion of the energy that exceeds the circuit power consumption of the tag.

\begin{figure}
	\centering
	\begin{minipage}[b]{0.45\textwidth}\centering
		\includegraphics[width=0.99\textwidth]{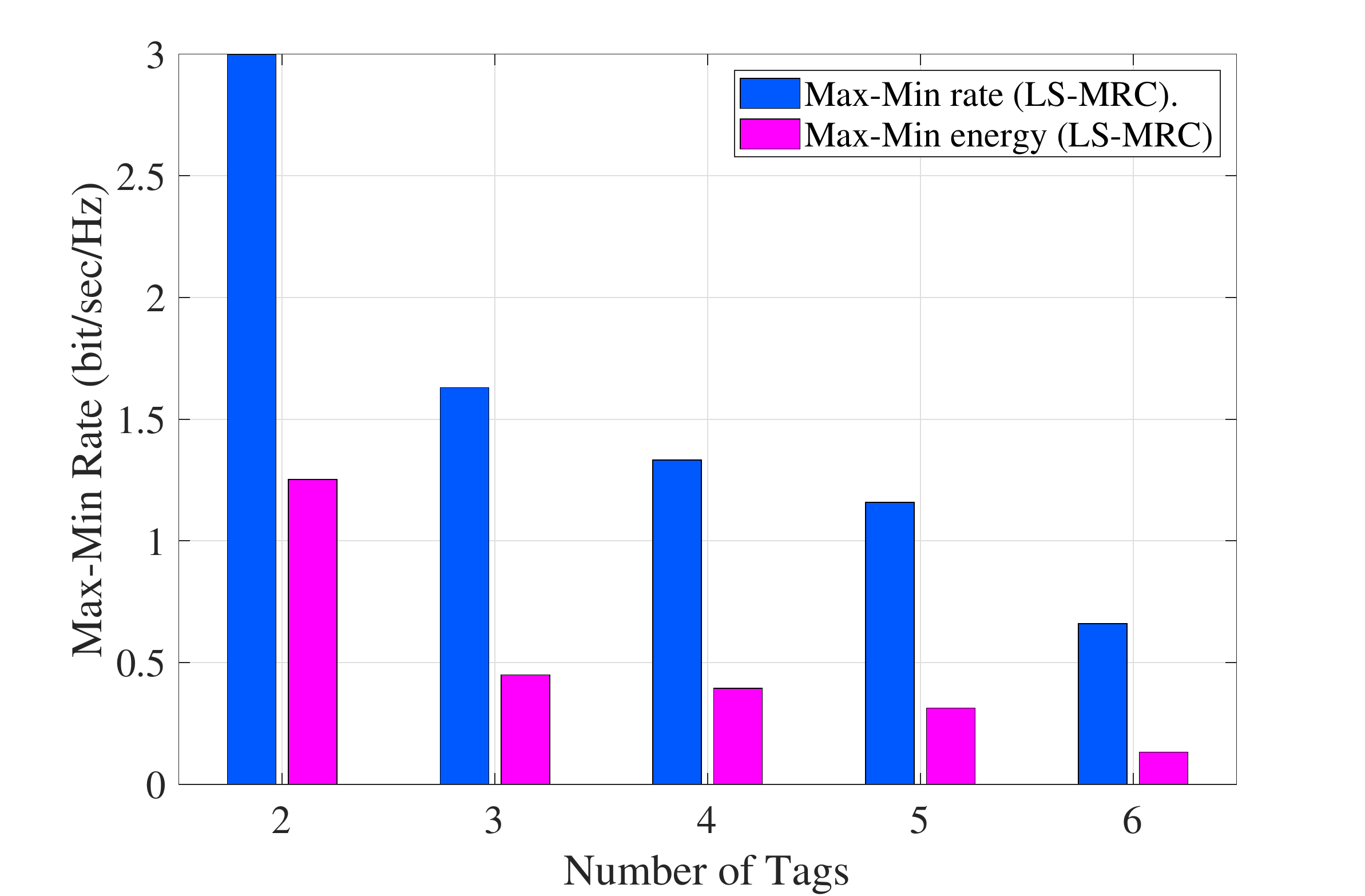}\vspace{-0.3cm}
		\caption{Maximal minimum achievable rate vs. $K$ for different optimization goals.} 
		\label{fig:multitag}
	\end{minipage}
	\begin{minipage}[b]{0.45\textwidth}\centering
		\includegraphics[width=0.99\textwidth]{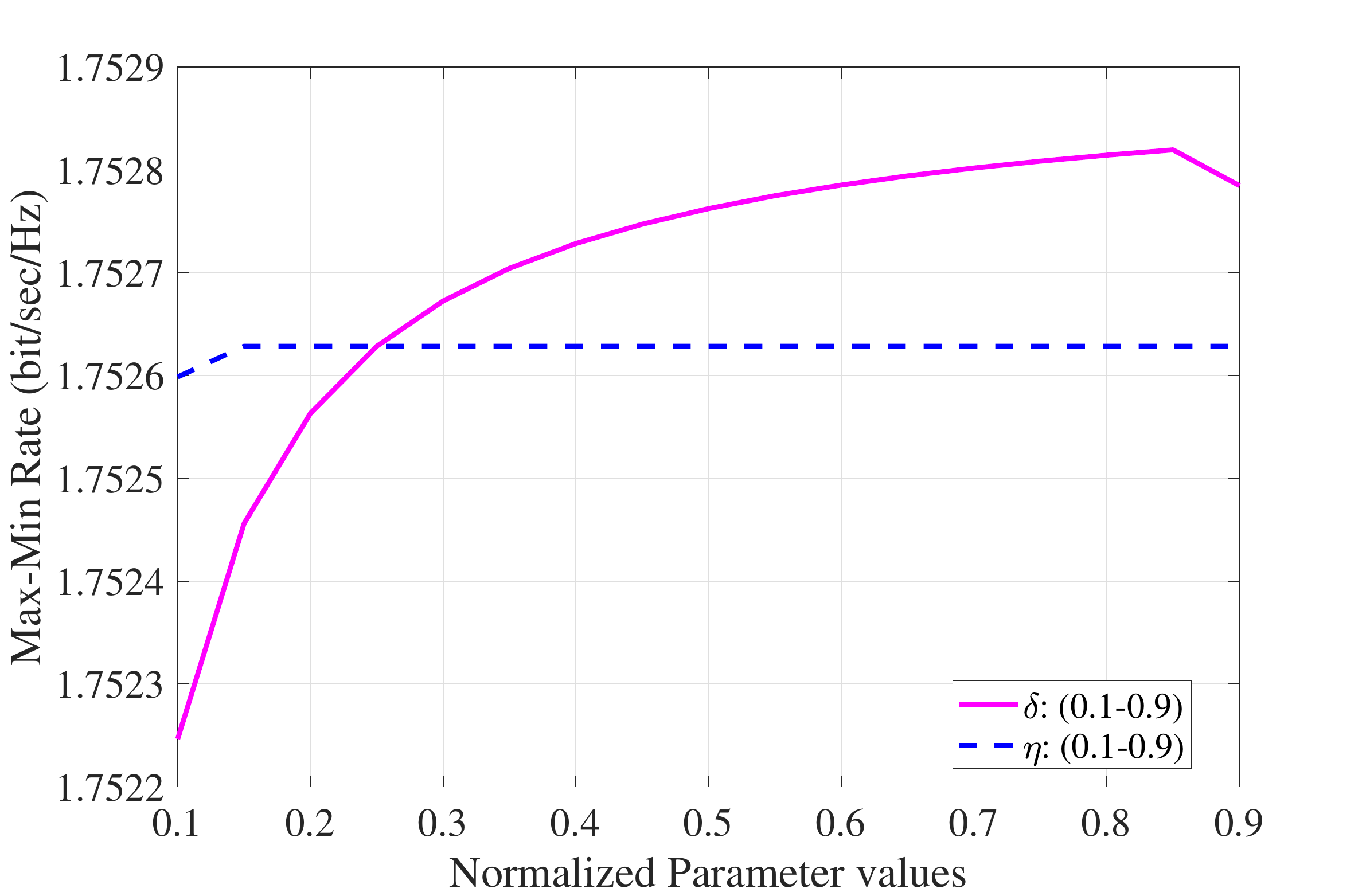}\vspace{-0.3cm}
		\caption{Maximal minimum achievable rate vs. different parameters like rectifier efficiency $\eta$, power reflection coefficient $\delta$.} 
		\label{fig:figurepara}
	\end{minipage}
\end{figure}


To demonstrate that our solution still performs well in a multi-user network, we plot the max-min rate under different number of tags in Fig.~\ref{fig:multitag}. The results show that our scheme always achieves a higher rate than the max-min energy-based scheme when numbers of tags are large. This is because our scheme can not only alleviate the double near-far effect by adjusting the beams but also makes use of the portion of the energy that exceeds the circuit power consumption.


Finally, we quantify the impact of key system parameters like rectifier efficiency $\eta$, power reflection coefficient $\delta$ on the max-min rate of our proposed scheme. The corresponding results plotted in Fig.~\ref{fig:figurepara} show that the max-min rate increases and then remains the same as $\eta$ increases and increases and then decreases as $\delta$ increases, respectively. This is because when the rectifier efficiency $\eta$ is very small, the circuit power consumption needs to be guaranteed by sacrificing the rate. When $\eta$ is large enough, the energy harvesting rate exceeds the circuit power consumption rate. Thus, the system can achieve the optimal max-min rate and remains the same with the increase of $\eta$. Besides, the rate first increases due to the reflected signal power increasing with the increase of power reflection coefficient $\delta$. When a certain critical point is reached, the remaining signal energy is not enough to meet the energy constraint. Thus, the max-min rate decreases to ensure that the energy constraint. These are consistent with our theoretical analysis.

\section{Research Extensions}\label{sec:Extensions}
In this section, we include a brief discussion on the non-linear energy harvesting model, imperfect self-interference cancellation, spatial multiplexing method for channel estimation and different fading models, which are worthing investigating in future work.

\subsection{Non-linear Energy Harvesting Model}
Since RF-based energy harvesting circuits that introduce various nonlinearities, stating the relationship between the output DC power of the energy harvester and its input RF power is difficult. There are two main different types of energy harvesting models, that is, the linear model and non-linear model of the energy harvester. Due to the benefit of being analytically easily tractable and good system performance in the low power regime, the linear energy harvesting model has been extensively studied~\cite{linear1,linear2,linear3}. Since the focus of this paper is the multi-user WPBC beamforming design, we use the linear energy harvesting model, which is driven by a tradeoff between accuracy and tractability, to bring some insightful results. The non-linear model is more accurately characterizes the rectenna behavior than the linear model in the practical observed from circuit simulation~\cite{8476597}. However, the complexity and intractability of the non-linear model hinder its good application and make it in infancy when it comes to multi-user WPBC design~\cite{8527670}. We tend to investigate the impact of the non-linear energy harvesting model on multi-user beamforming design for WPBC in future work.

\subsection{Imperfect Self-interference Cancellation}
This work focuses on the beamforming design for multiple tags, and we estimate the backscatter channels while assuming the perfect self-interference cancellation. However, in practice, perfect self-interference cancellation is difficult. To suppress the residual self-interferences, we can perform a pre-estimation of the unintended reflection link due to imperfect self-cancellations in the CE phase~\cite{2019Multi-Tag}. Specifically, denoting the residual self-interference channel by the matrix $ \textbf{H}_{R} \in{\mathbb{C}^{R\times M}} $ when the reader transmits the pilot sequence along with silent tags, we can first using the LS or MMSE estimators as proposed in Section~\ref{sec:estimator} to attain the estimation value of the residual self-interference channel as $\widehat{\textbf{H}}_{R}$. Next, we obtain the effective received signal at the reader after residual self-interference suppression. Finally, the backscatter channel can be estimated in the same way as discussed in Section~\ref{sec:estimator} with the effective received signal. In the information transfer phase, the received signal vector at the reader from $K$ tags can also be expressed as Eq.~\eqref{RSV_IT} when the residual self-interference exists. However, $ \boldsymbol{\mu}$ denotes contributions from both AWGN and residual self-interference at the reader with the variance for its entries being $\sigma_{\boldsymbol{\mu}}^2 $. Therefore, the SINR is actual and the beamforming design is based on the channel estimation after residual self-interference suppression.

\subsection{Spatial Multiplexing Method for Channel Estimation}
In this work with the aim of investigating a multi-user WPBC beamforming design, we first estimate the backscatter channel matrix $\textbf{H}_{k}$ of each tag by controlling tags to reflect the pilot signal one-by-one. Though this operation seems inefficient to estimate the backscatter channels, it can be easy to separate simultaneously received signals from all tags for subsequence analysis like to the orthogonal preamble designing in conventional MIMO communication. This is a spatial multiplexing method for each backscatter channel estimation. However, we would like to give some insight on simultaneous estimating the backscatter channel of $K$ tags in the channel estimation stage to reduce the overhead of channel estimation. Thus, the received signal vector from $K$ tags can be represented by $\textbf{y}^{\ast}=\textbf{H}^{b}\boldsymbol{\Psi} \left(\textbf{H}^{f}\right)^{T}\textbf{x}_{ref}+\textbf{N}$, where $\boldsymbol{\Psi}$, $\textbf{x}_{ref}$, and $\textbf{N}$ are the backscatter data, the pilot signal transmit from the reader, and the noise. If we can modulate the backscatter data as an identity matrix, the backscatter channel of $K$ tags is given by $\textbf{H}=\textbf{H}^{b}\left(\textbf{H}^{f}\right)^{T}$. The element $[\textbf{H}]_{mr}=\sum_{k=1}^{K}h_{mk}^{f}h_{kr}^{b}$, which is difficult to separate each backscatter channel directly like this work. This method will be requiring a new and dedicated investigation in future work.

\subsection{Different Fading Channels}
Due to the rich scattering in a complex environment of the large coverage range WPBC network, we consider that all the links suffer from the Rayleigh fading. However, there is a well-known fading channel that is good fit for the actual measured data, referred to as Nakagami-m. It includes the Rayleigh model as the special cases with $m = 1$ and is approximately equivalent to fading channel with LoS such as Rician channel when $m > 1$, respectively. Although our approach of energy beamforming is general in that it can also work well under any other channel model, the analysis depends on the distribution of the channel fading gains. Thus, our analysis cannot be directly used for any other fading channels. It requires brand new analysis to incorporate the Nakagami-m model into our work and thus deserve further study.

\section{Conclusion}\label{sec:conclusion}
In this paper, we investigate the energy beamforming via estimated BS-CSI to improve the performance of a multiuser WPBC system. To ensure uninterrupted communication, achievable rate and user fairness, we propose a beamforming scheme for power and information transfer in WPBC using the estimated BS-CSI. We use both LS and MMSE approaches to estimate backscatter channel matrix at the reader to perform energy beamforming. This estimation scheme is preferable in practical scenarios due to its affordable complexity and overhead. Additionally, we obtain the analytical expressions of the energy harvesting rate and the lower bound on the ergodic achievable rate to optimize resource allocation for maximizing the minimum rate for all backscatter tags. Our results indicate that the proposed beamforming scheme offers tremendous performance gains compared to the state-of-the-art energy beamforming schemes. We hope that our investigation on the proposed energy beamforming scheme can provide some implications for future designs.  

\vspace{-0.3cm}
\appendices

	\section{Proof of Lemma 1} 
	Let $\textbf{x}=\widehat{\textbf{h}}_{kr}, \textbf{y}=\textbf{h}_{k}^{f}$, we can note that $\textbf{x}$ and $\textbf{y}$ are jointly Gaussian distribution. 	
	\begin{equation}
	\textbf{z}=\begin{bmatrix}\textbf{x}  \\\textbf{y}   \end{bmatrix}=\begin{bmatrix}\widehat{\textbf{h}}_{kr}  \\\textbf{h}_{k}^{f}   \end{bmatrix}=\begin{bmatrix}\textbf{G}_{kr}\textbf{h}_{k}^{f}+\tilde{\textbf{e}}_{k}  \\\textbf{h}_{k}^{f}   \end{bmatrix}=\begin{bmatrix}\textbf{G}_{kr} & \textbf{I}_{M} \\\textbf{I}_{M} & \textbf{0}_{M} \end{bmatrix}\begin{bmatrix}\textbf{h}_{k}^{f}  \\\tilde{\textbf{e}}_{k}  \end{bmatrix}.
	\end{equation}
	Then we can obtain the mean
	\begin{equation}\label{mean}
	\begin{aligned}
	\mathbb E[\textbf{x}]=\mathbb E[\textbf{G}_{kr}\textbf{h}_{k}^{f}+\tilde{\textbf{e}}_{k}]=\textbf{0}, \quad \mathbb E[\textbf{y}]=\textbf{0}.
	\end{aligned}
	\end{equation}
	Since $\textbf{h}_{k}^{f}$ is independent of $\tilde{\textbf{e}}_{k}$, we can obtain the covariance
	\begin{equation}\label{covariance}
	\begin{aligned}
	\textbf{C}_{\textbf{x}\textbf{x}}&=\mathbb E[(\textbf{x}-\mathbb E(\textbf{x})) (\textbf{x}-\mathbb E(\textbf{x}))^{H}]=\mathbb E[\textbf{x}\textbf{x}^{H}]\\&=\textbf{G}_{kr}\mathbb E[\textbf{h}_{k}^{f}\left(\textbf{h}_{k}^{f}\right)^{H}]\textbf{G}_{kr}^{H}+\mathbb E[\tilde{\textbf{e}}_{k}\tilde{\textbf{e}}_{k}^{H}]\\&=\textbf{G}_{kr}\textbf{C}_{\textbf{h}_{k}^{f}}\textbf{G}_{kr}^{H}+ \textbf{C}_{\tilde{\textbf{e}}_{k}}.
	\end{aligned}
	\end{equation} 
	In addition, the cross-covariance matrix can be expressed as
	\begin{equation}\label{cross-covariance}
	\begin{aligned}
	\textbf{C}_{\textbf{y}\textbf{x}}&=\mathbb E[(\textbf{y}-\mathbb E(\textbf{y})) (\textbf{x}-\mathbb E(\textbf{x}))^{H}]=\mathbb E[\textbf{h}_{k}^{f}(\textbf{G}_{kr}\textbf{h}_{k}^{f}+\tilde{\textbf{e}}_{k})^{H}]=\textbf{C}_{\textbf{h}_{k}^{f}}\textbf{G}_{kr}^{H}.
	\end{aligned}
	\end{equation} 
	According to the theorem of conditional PDF of multivariate Gaussian~\cite{Kay1993fundamentals}, the conditional PDF of $\textbf{y}|\textbf{x}$ is also Gaussian and
	\begin{equation}\label{posterior}
	\begin{aligned}
	\mathbb E[\textbf{y}|\textbf{x}]&=\mathbb E[\textbf{y}]+\textbf{C}_{\textbf{y}\textbf{x}}\textbf{C}_{\textbf{x}\textbf{x}}^{-1}(\textbf{x}-\mathbb E[\textbf{x}]),\\\textbf{C}_{\textbf{y}|\textbf{x}}&=\textbf{C}_{\textbf{y}\textbf{y}}-\textbf{C}_{\textbf{y}\textbf{x}}\textbf{C}_{\textbf{x}\textbf{x}}^{-1}\textbf{C}_{\textbf{x}\textbf{y}}.
	\end{aligned}
	\end{equation}
	Then, we can substitute \eqref{mean}, \eqref{covariance} and \eqref{cross-covariance} into \eqref{posterior}. Here we complete the proof.

	\section{Proof of Proposition 1} 
	Since $h_{kr}^{b} \sim \mathcal{CN}\left( 0, \beta_{k }\right)$, define the random variable $f=|h_{kr}^{b}|^2$, it can be easily checked that $f$ follows exponential distribution. Then, we can obtain the expectation in~\eqref{incident signal power} as follows.
	\begin{align}\label{expectation}
	\phi\left(h_{kr}^{b}\right)&=\begin{cases} \mathbb E_{h_{kr}^{b}}\left\lbrace\frac{1}{\frac{|h_{kr}^{b}|^2\beta_{k}\alpha p_{ce}  \delta_{k}}{K\sigma^2}+1}\right\rbrace, & for\quad LS,\\\mathbb E_{h_{kr}^{b}}\left\lbrace\frac{1}{\frac{|h_{kr}^{b}|^2(K\sigma^2+\beta_{k}^2\alpha p_{ce}  \delta_{k})}{\beta_{k}K\sigma^2}+1}\right\rbrace, & for\quad MMSE,\end{cases}\\
	&=\begin{cases}\frac{K\sigma^2}{\beta_{k}^2 \alpha p_{ce} \delta_{k}} exp(\frac{K\sigma^2}{\beta_{k}^2 \alpha p_{ce} \delta_{k}}) \Gamma(0, \frac{K\sigma^2}{\beta_{k}^2 \alpha p_{ce} \delta_{k}}), & for\quad LS,\\\frac{K\sigma^2}{K\sigma^2+\beta_{k}^2\alpha p_{ce} \delta_{k}} exp(\frac{K\sigma^2}{K\sigma^2+\beta_{k}^2\alpha p_{ce} \delta_{k}}) \Gamma(0, \frac{K\sigma^2}{K\sigma^2+\beta_{k}^2\alpha p_{ce} \delta_{k}}), & for\quad MMSE,\end{cases}
	\end{align}
	 where $\Gamma(0,t)\triangleq\int_{t}^{\infty}u^{-1}\exp(-u)du$. Thus, the lower bound and upper bound in Proposition~\ref{boundary} can be obtained according to the bound of Gamma function~\cite{abramowitz1964handbook}.

	\section{Proof of Proposition 2} 
    Define $\tilde{\psi}_{i} \triangleq \frac{\widehat{\textbf{h}}_{mk}^{H} \widehat{\textbf{h}}_{mi}}{\|\widehat{\textbf{h}}_{mk}\|}$ and $\tilde{\psi}_{i} \sim \mathcal{CN}\left(0,|h_{mi}^{f}|^2\left( \beta_{i}+\sigma_{\epsilon,mi}^2(h_{mi}^{f})\right) \right)$, where $\sigma_{\epsilon, mi}^2(h_{mi}^f)=\frac{\sigma_{\tilde{e},i}^2}{|h_{mi}^{f}|^2}$. Since $h_{mk}^{f}$ and $h_{mi}^{f}$ are independent, we can use the property of expectation to rewrite the expectation in (26) as
    	\begin{equation}\label{MRC_EX}
    	\begin{aligned}
    	&\mathbb{E} \left\lbrace \frac{\sum_{i=1,i\neq k}^{K}  p_{i} \frac{1}{|h_{mi}^{f}|^2} |\tilde{\psi}_{i}|^2  + \sum_{i=1}^{K}  p_{i}  \sigma_{\epsilon,mi}^2(h_{mi}^{f})+ \sigma^2}  {p_{k} \frac{1}{|h_{mk}^{f}|^2} \|\widehat{\textbf{h}}_{mk}\|^2 }\right\rbrace\\
    	= &\mathbb{E} \left\lbrace \frac{\sum_{i=1,i\neq k}^{K}  p_{i} \frac{1}{|h_{mi}^{f}|^2} |\tilde{\psi}_{i}|^2  + \sum_{i=1,i\neq k}^{K}  p_{i}  \sigma_{\epsilon,mi}^2(h_{mi}^{f}) + p_{k}  \sigma_{\epsilon,mk}^2(h_{mk}^{f}) + \sigma^2}  {p_{k} \frac{1}{|h_{mk}^{f}|^2} \|\widehat{\textbf{h}}_{mk}\|^2 }\right\rbrace\\
    	= &\mathbb{E} \left\lbrace \frac{\sum_{i=1,i\neq k}^{K}  p_{i} \frac{1}{|h_{mi}^{f}|^2} |\tilde{\psi}_{i}|^2  + \sum_{i=1,i\neq k}^{K}  p_{i}  \sigma_{\epsilon,mi}^2(h_{mi}^{f})  + \sigma^2}  {p_{k} \frac{1}{|h_{mk}^{f}|^2} \|\widehat{\textbf{h}}_{mk}\|^2 }\right\rbrace + \mathbb{E} \left\lbrace \frac{ \sigma_{\epsilon,mk}^2(h_{mk}^{f})}{\frac{1}{|h_{mk}^{f}|^2} \|\widehat{\textbf{h}}_{mk}\|^2} \right\rbrace\\
    	= & \bigg( \sum_{i=1,i\neq k}^{K} p_{i}\beta_{i}  + 2\sum_{i=1,i\neq k}^{K} p_{i} \sigma_{\tilde{e},i}^2 \mathbb{E}_{h_{mi}^{f}} \left\lbrace\frac{1}{|h_{mi}^{f}|^2} \right\rbrace + \sigma^2 \bigg)
    	\mathbb{E} \left\lbrace \frac{1}{ p_{k} \frac{1}{|h_{mk}^{f}|^2} \|\widehat{\textbf{h}}_{mk}\|^2} \right\rbrace + \mathbb{E} \left\lbrace \frac{\sigma_{\tilde{e},k}^2}{\|\widehat{\textbf{h}}_{mk}\|^2} \right\rbrace.\\
    	\end{aligned}
    	\end{equation}

Define the random variable $\tilde{x}_{i} \triangleq |h_{mi}^{f}|^2 $, which can be easily checked that $\tilde{x}_{i} \sim \exp\left( \frac{1}{\beta_{i}}\right)$. We further define the function $g(\tilde{x}_{i}) \triangleq \frac{1}{\tilde{x}_{i}}$, where $\tilde{x}_{i}\in(0,\infty)$, and $\tau>0$ as the minimum of $\tilde{x}_{i}$.
    Then, we use the expectation of the random variable function and the original definition of expectation to rewrite the expectation as
    \begin{equation}\label{ex}
    \begin{aligned}
    \mathbb{E}_{h_{mi}^{f}} \left\lbrace\frac{1}{|h_{mi}^{f}|^2}\right\rbrace  = \int_{\tau}^{\infty} \frac{1}{\beta_{i} \tilde{x}_{i}} e^{-\frac{\tilde{x}_{i}}{\beta_{i}}} d\tilde{x}_{i}= \frac{\Gamma(0,\frac{\tau}{\beta_{i}})}{\beta_{i}},
    \end{aligned} 
    \end{equation}
    where $\Gamma(0,\frac{\tau}{\beta_{i}}) \triangleq \int_{\frac{\tau}{\beta_{i}}}^{\infty} t^{-1} e^{-t} dt$.
	
	Since $\widehat{\textbf{h}}_{mk}|h_{mk}^{f} \sim \mathcal{CN}\left(\textbf{0}_{R},|h_{mk}^{f}|^2\left( \beta_{k}+\sigma_{\epsilon,mk}^2(h_{mk}^{f})\right) \textbf{I}_{R}\right) $, $Z_{mk} \triangleq \frac{2}{|h_{mk}^{f}|^2\left( \beta_{k}+\sigma_{\epsilon,mk}^2(h_{mk}^{f})\right)} \widehat{\textbf{h}}_{mk}^{H} \widehat{\textbf{h}}_{mk}$ follows central chi-square distribution with $2R$ degree of freedom. Thus, the $\frac{1}{Z_{mk}} \sim$ Inv-$\chi^{2}(2R)$, and we have $\mathbb{E} \left\lbrace\frac{1}{Z_{mk}}\right\rbrace = \frac{1}{2(R - 1)}$. Then
	\begin{equation}\label{ex2}
	\begin{aligned}
	\mathbb{E} \left\lbrace \frac{1}{ p_{k} \frac{1}{|h_{mk}^{f}|^2} \|\widehat{\textbf{h}}_{mk}\|^2} \right\rbrace  &= \mathbb{E} \left\lbrace \frac{2}{ p_{k} \left( \beta_{k} +\sigma_{\epsilon,mk}^2(h_{mk}^{f}) \right) Z_{mk}} \right\rbrace\\
	& = \mathbb{E}_{h_{mk}^{f}} \left\lbrace \frac{1}{ p_{k} (R-1) \left( \beta_{k} +\sigma_{\epsilon,mk}^2(h_{mk}^{f}) \right) } \right\rbrace\\
	& = \frac{1}{p_{k} (R-1)\beta_{k}} \mathbb{E}_{h_{mk}^{f}} \left\lbrace \frac{1}{1 + \frac{\sigma_{\tilde{e},k}^2}{\beta_{k}|h_{mk}^{f}|^2 }} \right\rbrace\\
	& = \frac{1}{p_{k} (R-1)\beta_{k}} \left( 1 - \mathbb{E}_{h_{mk}^{f}} \left\lbrace \frac{1}{\frac{\beta_{k}|h_{mk}^{f}|^2}{\sigma_{\tilde{e},k}^2}+1} \right\rbrace\right).\\                     
	\end{aligned} 
	\end{equation}	
		\begin{equation}\label{ex23}
		\begin{aligned}
		\mathbb{E} \left\lbrace \frac{\sigma_{\tilde{e},k}^2}{\|\widehat{\textbf{h}}_{mk}\|^2} \right\rbrace & = \mathbb{E} \left\lbrace \frac{2\sigma_{\tilde{e},k}^2}{|h_{mk}^{f}|^2\left( \beta_{k} +\sigma_{\epsilon,mk}^2(h_{mk}^{f}) \right) Z_{mk}} \right\rbrace\\
		& = \mathbb{E}_{h_{mk}^{f}} \left\lbrace \frac{\sigma_{\tilde{e},k}^2}{ (R-1) |h_{mk}^{f}|^2 \left( \beta_{k} +\sigma_{\epsilon,mk}^2(h_{mk}^{f}) \right) } \right\rbrace\\
		& = \frac{1}{(R-1)} \mathbb{E}_{h_{mk}^{f}} \left\lbrace \frac{1}{\frac{\beta_{k}|h_{mk}^{f}|^2}{\sigma_{\tilde{e},k}^2}+1} \right\rbrace.\\
		\end{aligned} 
		\end{equation}
		Next, substituting~\eqref{ex},~\eqref{ex2} and~\eqref{ex23} into~\eqref{MRC_EX}, we can attain that 
		\begin{equation}\label{MRC_EX1}
		\begin{aligned}
		& \bigg( \sum_{i=1,i\neq k}^{K} p_{i}\beta_{i}  + 2\sum_{i=1,i\neq k}^{K} p_{i} \sigma_{\tilde{e},i}^2 \mathbb{E}_{h_{mi}^{f}} \left\lbrace\frac{1}{|h_{mi}^{f}|^2} \right\rbrace + \sigma^2 \bigg)
		\mathbb{E} \left\lbrace \frac{1}{ p_{k} \frac{1}{|h_{mk}^{f}|^2} \|\widehat{\textbf{h}}_{mk}\|^2} \right\rbrace + \mathbb{E} \left\lbrace \frac{\sigma_{\tilde{e},k}^2}{\|\widehat{\textbf{h}}_{mk}\|^2} \right\rbrace\\
		=& \bigg( \sum_{i=1,i\neq k}^{K} p_{i}\beta_{i}  + 2\sum_{i=1,i\neq k}^{K} p_{i} \sigma_{\tilde{e},i}^2 \frac{\Gamma(0,\frac{\tau}{\beta_{i}})}{\beta_{i}} + \sigma^2 \bigg)
		\frac{ \left(1-\phi\left(h_{mk}^{f}\right)\right)}{p_{k} (R-1)\beta_{k}} + \frac{ \phi\left(h_{mk}^{f}\right)}{(R-1)}\\
		=& \frac{\bigg( \sum_{i=1,i\neq k}^{K} p_{i}\beta_{i}  + 2\sum_{i=1,i\neq k}^{K} p_{i} \sigma_{\tilde{e},i}^2 \frac{\Gamma(0,\frac{\tau}{\beta_{i}})}{\beta_{i}} + \sigma^2 \bigg) \left(1-\phi\left(h_{mk}^{f}\right)\right) + p_{k}\beta_{k} \phi\left(h_{mk}^{f}\right)}{p_{k} (R-1)\beta_{k}}.
		\end{aligned}
		\end{equation}
		
		Then, we can substitute~\eqref{MRC_EX1} into~(26). Here we complete the proof.
	
	\section{Proof of Proposition 3} 
	With ZF detector, $\textbf{Q}^{H}= \left(\widehat{\textbf{H}}_{m}^{H}\widehat{\textbf{H}}_{m}\right)^{-1}\widehat{\textbf{H}}_{m}$. Therefore,
	$\textbf{q}_{k}^{H} \widehat{\textbf{h}}_{i}^{b}= \begin{cases}\frac{1}{h_{mk}^{f}}, & if\quad k=i\\0, & otherwise\end{cases}$.
	From~\eqref{lower boound}, the achievable rate of tag $k$ can be expressed as
	\begin{equation}\label{rate_zf}
	\begin{aligned}
	\tilde{R}^{ZF}_{k} = \log_{2}\left(1 + \left(\mathbb{E} \left\lbrace \frac{\left(\sum_{i=1}^{K}  p_{i}  \sigma_{\epsilon,mi}^2(h_{mi}^{f})+ \sigma^2\right) \left[\left(\widehat{\textbf{H}}_{m}^{H}\widehat{\textbf{H}}_{m}\right)^{-1}\right]_{kk}}{\frac{p_{k}}{|h_{mk}^{f}|^2}} \right\rbrace \right)^{-1}\right).
	\end{aligned}
	\end{equation}
	We can know that $\widehat{\textbf{h}}_{mk}|h_{mk}^{f} \sim \mathcal{CN}\left(\textbf{0}_{R},|h_{mk}^{f}|^2( \beta_{k}+\sigma_{\epsilon,mk}^2(h_{mk}^{f})) \textbf{I}_{R}\right) $ from the derivation results in Section~\ref{sec:model}-B. Then, using the property of Wishart Matrix~\cite{Tulino2004Random}, we can obtain 
	\begin{equation}
	\begin{aligned}
	\mathbb{E} \left\lbrace \left[\left(\widehat{\textbf{H}}_{m}^{H}\widehat{\textbf{H}}_{m}\right)^{-1} \right]_{kk} \right\rbrace&=\frac{1}{K}\mathbb{E} \left\lbrace tr \left[\left(\widehat{\textbf{H}}_{m}^{H}\widehat{\textbf{H}}_{m}\right)^{-1} \right] \right\rbrace\\
	&=\frac{1}{\left(R-K\right) |h_{mk}^{f}|^2( \beta_{k}+\sigma_{\epsilon,mk}^2(h_{mk}^{f}))}.
	\end{aligned}
	\end{equation}
	Then, we have	
	\begin{equation}\label{ex3}
	\begin{aligned}
	&\mathbb{E} \left\lbrace \frac{\left(\sum_{i=1}^{K}  p_{i}  \sigma_{\epsilon,mi}^2(h_{mi}^{f})+ \sigma^2\right) \left[\left(\widehat{\textbf{H}}_{m}^{H}\widehat{\textbf{H}}_{m}\right)^{-1}\right]_{kk}}{\frac{p_{k}}{|h_{mk}^{f}|^2}} \right\rbrace\\
	=&\mathbb{E} \left\lbrace \frac{\sum_{i=1,i\neq k}^{K}  p_{i}  \sigma_{\epsilon,mi}^2(h_{mi}^{f})+ \sigma^2 }{p_{k}\left(R-K\right) ( \beta_{k}+\sigma_{\epsilon,mk}^2(h_{mk}^{f}))} \right\rbrace + \mathbb{E} \left\lbrace \frac{  \sigma_{\epsilon,mk}^2(h_{mk}^{f})}{\left(R-K\right) ( \beta_{k}+\sigma_{\epsilon,mk}^2(h_{mk}^{f}))} \right\rbrace
	\end{aligned}
	\end{equation}
	\begin{equation*}
    \begin{aligned}
    =&\frac{1}{p_{k} (R-K)\beta_{k}} \left(\sum_{i=1,i\neq k}^{K}  p_{i} \sigma_{\tilde{e},i}^2\mathbb{E}_{h_{mi}^{f}} \left\lbrace\frac{1}{|h_{mi}^{f}|^2}\right\rbrace + \sigma^2 \right)  \left( 1 - \mathbb{E}_{h_{mk}^{f}} \left\lbrace \frac{1}{\frac{\beta_{k}|h_{mk}^{f}|^2}{\sigma_{\tilde{e},k}^2}+1} \right\rbrace\right)\\
     &+\frac{1}{(R-K)} \mathbb{E}_{h_{mk}^{f}} \left\lbrace \frac{1}{\frac{\beta_{k}|h_{mk}^{f}|^2}{\sigma_{\tilde{e},k}^2}+1} \right\rbrace.
    \end{aligned}
    \end{equation*}

Substituting~\eqref{ex} and~\eqref{ex3} into~\eqref{rate_zf}, Proposition~\ref{zf_rate} is proved.


\ifCLASSOPTIONcaptionsoff
  \newpage
\fi



\bibliographystyle{IEEEtranTCOM}
\bibliography{IEEEabrv,./WPBC}
%
%
%
%

%




\end{document}